\documentclass[12pt,preprint]{aastex}
\usepackage[T1]{fontenc} 
\usepackage[english]{babel}    
\RequirePackage{amsfonts}
\RequirePackage{amssymb}

\newcommand{\scname}[1]{\textsc{#1}}

\newcommand{\snap}{\scname{SNAP}} 
\newcommand{\phoenix}{\texttt{PHOENIX}}

\newcommand{\kms}{\ensuremath{\textrm{km}~\textrm{s}^{-1}}}

\newcommand{\supernovae}{supernov\ae} 
\newcommand{\Supernovae}{Supernov\ae} 
\newcommand{\SNIa}{{SN~Ia}} 
\newcommand{\SNeIa}{{SNe~Ia}}

\newcommand{\SiII}{Si~\scname{ii}}

\newcommand{\FeII}{Fe~\scname{ii}}

\newcommand{\CaII}{Ca~\scname{ii}}

\newcommand{\SiIIred}{\SiII $\lambda 6355$}
\newcommand{\SiIIblue}{\SiII $\lambda 5972$}

\newcommand{\RCaS}{\ensuremath{\Re_{CaS}}}

\newcommand{\RSiSu}{\ensuremath{\Re_{SiS}}}
\newcommand{\RSiSuS}{\ensuremath{\Re_{SiSS}}}

\newcommand{\nomw}{W7}

\newcommand{\snd}{\ensuremath{2^{\textrm{nd}}}} 
 
\def\xth#1{{#1\ensuremath{^{\textrm{th}}}}}

\RequirePackage{amsfonts}
\RequirePackage{amssymb}

\def\ifundefined#1{\expandafter\ifx\csname#1\endcsname\relax}

\def\la{\mathrel{\hbox{\rlap{\hbox{\lower4pt\hbox{$\sim$}}}\hbox{$<$}}}}
\def\ga{\mathrel{\hbox{\rlap{\hbox{\lower4pt\hbox{$\sim$}}}\hbox{$>$}}}}

\newcommand{\be}{\begin{eqnarray}}
\newcommand{\ee}{\end{eqnarray}}

\ifundefined{ensuremath}\def\ensuremath#1{\relax\ifmmode{#1}}
\else${#1}$\fi\else\relax\fi
\ifundefined{nuc}\def\nuc#1#2{\relax\ifmmode{}^{#1}{\protect\textrm{#2}}
\else${}^{#1}$#2\fi}\else\relax\fi

\newcommand{\etal}{et al.}

\newcommand{\kmps}{\ensuremath{\mbox{km}~\mbox{s}^{-1}}}

\newcommand{\msol}{\ensuremath{{\mbox{M}_\odot}}}

\def\Teff{\ensuremath{T_{\mbox{model}}}}

\newcommand{\RSi}{$\Re_{Si}$}
\newcommand{\RCa}{$\Re_{Ca}$}

\newcommand{\phx}{\texttt{PHOENIX}}

\graphicspath{{./}}
\DeclareGraphicsExtensions{{.eps},{.ps}}

\received{} 
\accepted{} 
\journalid{}{} 
\articleid{}{} 
\shortauthors{Bongard, S. et~al.}
\shorttitle{Type Ia Supernova Spectral Line Ratios}

\begin{document}

\title{Type Ia Supernova Spectral Line Ratios as Luminosity Indicators}

\author{ 
  Sebastien Bongard,\altaffilmark{1,2,4}\email{sbongard@lbl.gov}
  E.~Baron,\altaffilmark{2,3}\email{baron@nhn.ou.edu}
  G.~Smadja,\altaffilmark{4}\email{smadja@in2p3.fr}  David
Branch,\altaffilmark{2}\email{branch@nhn.ou.edu}  and  Peter
H.~Hauschildt\altaffilmark{5}\email{yeti@hs.uni-hamburg.de}
 }

\altaffiltext{1}{Physics Division, Lawrence Berkeley
  National Laboratory, MS 50R-5008, 1 Cyclotron Rd, Berkeley, CA
  94720 USA}

\altaffiltext{2}{Department of Physics and Astronomy, University of
Oklahoma, 440 West Brooks, Rm.~100, Norman, OK 73019-2061, USA}

\altaffiltext{3}{Computational Research Division, Lawrence Berkeley
  National Laboratory, MS 50F-1650, 1 Cyclotron Rd, Berkeley, CA
  94720 USA}

\altaffiltext{4}{Institute de Physique Nucl\'eaire Lyon, B\^atiment Paul Dirac
Universit\'e Claude Bernard Lyon-1
Domaine scientifique de la Doua
4, rue Enrico Fermi
69622 Villeurbanne cedex, France}

\altaffiltext{5}{Hamburger Sternwarte, Gojenbergsweg 112,
21029 Hamburg, Germany}

\begin{abstract}
  Type Ia supernovae have played a crucial role in the discovery of
  the dark energy, via the measurement of their light curves and the
  determination of the peak brightness via fitting templates to the
  observed lightcurve shape. Two spectroscopic indicators are also
  known to be well correlated with peak luminosity. Since the
  spectroscopic luminosity indicators are obtained directly from
  observed spectra, 
  they will have different systematic errors than do measurements
  using photometry. Additionally, these spectroscopic indicators may
  be useful for studies of effects of evolution or age of the SNe~Ia
  progenitor population. We present several new variants of such
  spectroscopic indicators which are easy to automate and which
  minimize the effects of noise. We show that these spectroscopic
  indicators can be measured by proposed JDEM missions such as \snap\
  and JEDI.
\end{abstract}


\section{Introduction}

Type Ia supernovae are now recognized as one of the most important
cosmological probes. 
Astronomical interest has been focused on Type Ia supernovae
since it was recognized long ago \citep{wilson39,kowal68} that they
are good ``standard candles'' and hence are useful cosmological
probes. Their use as cosmological beacons led to the
discovery of the ``dark 
energy'' \citep{riess_scoop98,garnetal98,perletal99} and they are an
integral part of plans to further characterize the nature of the dark
energy equation of state with a wide field satellite probe, the
``Joint Dark Energy Mission'' (JDEM) as well as ground-based studies.
The reliability of SNe Ia as distance indicators improved
significantly with the realization that the luminosity at peak was
correlated with the width of the light curve \citep{philm15} and hence
that SNe~Ia are correctable candles in much the same way that Cepheids
are \citep{philetal99,goldhetal01,rpk95}. To date, cosmology with Type
Ia \supernovae has used photometry exclusively over spectroscopy, except that a
spectrum is required for supernova type confirmation and can be used
for the redshift determination, the luminosity at peak is always
determined photometrically (by matching lightcurve
templates). However, there exist spectroscopic luminosity indicators
first discussed by \citet{nugseq95}.  \citet{garn99by04} showed that
\RSi\ \citep{nugseq95} is well-correlated with $\Delta m_{15}$. Of
course, the value of $M_B$ at maximum light is also well correlated
with $\Delta m_{15}$ \citep{philetal99,philm15}. Thus, the
spectroscopic luminosity indicator should be well correlated with the
peak luminosity, making it a useful secondary luminosity indicator for
cosmological studies. Even though it is correlated with peak
luminosity via the photometric indicator $\Delta m_{15}$, it is still
useful as an independent check because it will have different
systematic errors than light curve shape fitting via templates. 

We study the correlation of the two spectroscopic luminosity
indicators \RSi\ and \RCa\  with luminosity, both for observed
\supernovae and for synthetic spectra of the parameterized explosion
model W7 \citep{nomw7} calculated using \phx
\citep{hbjcam99,hbmathgesel04} in LTE. We define new, more robust
spectroscopic luminosity indicators and we show that these indicators
can be used with good accuracy via proposed JDEM missions such as SNAP
\citep{snap_detf_wp05} and JEDI \citep{jedi_detf_wp05}.

\section{Spectral Indices \RSi\ \& \RCa }
\label{sec:rsi--rca}

While most Type Ia \supernovae fall into the category of ``Branch
normal'' \citep{bfn93} with $B-V$ colors at peak close to 0, there are
known to be a class of bright ``1991T-like'' \supernovae that have
weak \SiII\ lines 
at maximum, but which develop characteristic \SiII\ lines shortly after
maximum light. Additionally there is a class of dim ``1991bg-like''
\supernovae, which are quite red and are thought to produce very low
amounts of nickel, $M_{\mbox{Ni}} \sim 0.1$~\msol\ as opposed to
$M_\textrm{Ni} \sim 0.6$~\msol\ for Branch normals \citep{mazz91bg97}.

\citet{nugseq95} showed that the diversity in the peak luminosity from
the dim 1991bg-likes, to the bright 1991T-likes could be interpreted
in terms of a single parameter \Teff in synthetic spectral models,
with the dim, low nickel mass 1991bg-likes being cool and the bright
1991T-likes being hot. Furthermore, \citet{nugseq95} found that the
ratio of the depth of the characteristic defining feature of SNe Ia,
\SiIIred, to that of another silicon line \SiIIblue\ is correlated with
the spectral sequence and hence with the peak luminosity. The obvious
interpretation is that this is a sequence in nickel mass produced,
although that interpretation is not completely proven. \citet{nugseq95}
also defined another spectroscopic indicator using the ratio of two
peaks near the Ca II H+K feature.

\begin{figure}
  \includegraphics[width = 0.8\textwidth, clip]{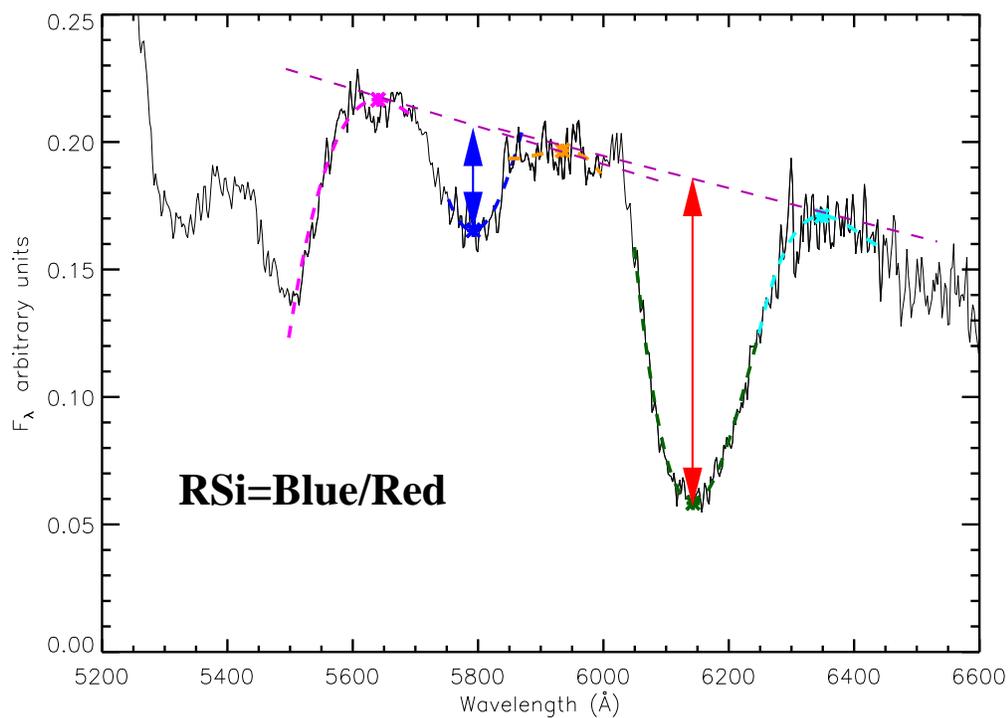}
  \caption[\RSi]{The definition of the method used for determining
    \RSi. The observed spectrum is that of SN~1992A 3 days after
    maximum and the 
    dashed lines indicate the polynomial fits to determine the maxima
    and minima. The arrows indicate the depths
    $d_\textrm{red}$ and $d_\textrm{blue}$ used to compute
    \RSi.}
  \label{fig:RSiDef}
\end{figure}

\begin{figure}
  \includegraphics[width = 0.8\textwidth, clip]{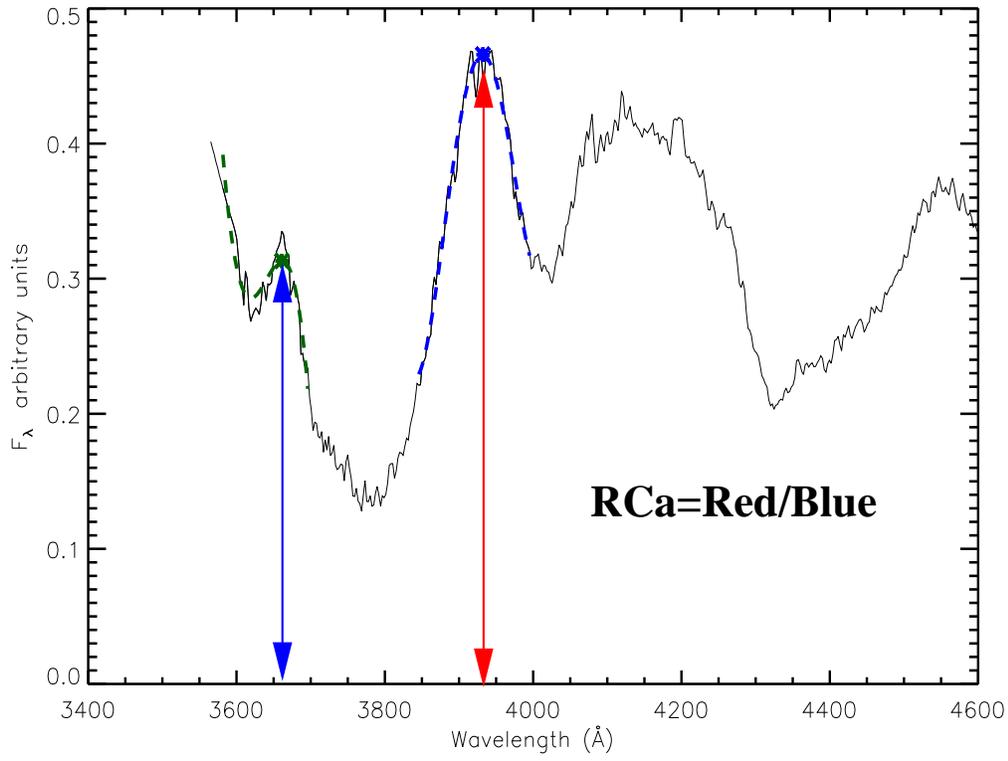}
  \caption[\RCa\  definition]{The definition of \RCa. The dashed lines
    indicate the polynomial fits used to determine the maxima, which
    are indicated by the arrows.
}
  \label{fig:RCaDef}
\end{figure}

\begin{deluxetable}{lcc}
\tablecolumns{3}
\tablewidth{0pc}
\tablecaption{\label{tab:rsi}\RSi\ definition zones}
\tablehead{\colhead{Feature} &
\colhead{Lower Wavelength (\AA)}    &   \colhead{Upper Wavelength (\AA)}}
\startdata
First maximum&5500 & 5700 \\
\SiIIblue\ minimum& Variable & Variable\\
Second maximum&5850 & 6000 \\
\SiIIred\ minimum& 6050& 6250 \\
First maximum&6200 & 6450 \\
\enddata
\end{deluxetable}

\subsection{Definition of \RSi}
\label{sec:rsi-1}

\RSi\ is named for the characteristic \SiII\ trough found
at $\approx 6100$\AA\ and usually associated with the absorption  
trough of the \SiIIred\ \SiII\ line P-Cygni feature.

Fig.~\ref{fig:RSiDef}, illustrates the definition of \RSi, using the
observed spectrum of SN~1992A 3 days after maximum. The dashed lines are the
polynomial fits used to determine the maxima found in the wavelength zones
defined in Table~\ref{tab:rsi}. The two depths are
indicated by arrows.

The \SiIIred\ minimum band is centered on the
\SiIIred\ line blueshifted approximately
$10,000$~\kms (Due to 
  homologous  expansion, it is customary to use velocities
  interchangeably with 
  wavelength shifts when  analyzing \supernovae spectra.). The value
  of 10,000~\kms
corresponds to the average velocity at which the \SiII\ line is
observed to form. The $\approx 200$\AA~ wide wavelength zone where we
search for the second minimum is noted as ``variable'' because it is
centered on the \SiIIblue\ line blueshifted by nearly the same actual velocity
as the \SiIIred\ minima. This to ensure that the right trough is
found, which is
assumed to be at least partly created by the \SiIIblue\ \SiII\ line,
thus it should form in the same physical region.

\RSi\ is defined as the ratio of the vertical distances between
each minima and the reference lines drawn between two consecutive
maxima $d_\textrm{blue}$ and $d_\textrm{red}$, as shown in
Fig.~\ref{fig:RSiDef},   

\begin{equation}
  \label{eq:47}
  \textrm{\RSi}=\frac{d_\textrm{blue}}{d_\textrm{red}}.
\end{equation}

This definition differs slightly from the one of \cite{nugseq95},
where tangents to the spectrum were manually selected instead of
the reference lines between two consecutive maxima. These two methods
mainly differ when the zones searched for maxima only display
inflection points, but our method was easier to automate.

Finally, because it is defined as the distance between points of the
spectrum and references linearly coupling points of the spectrum, \RSi
is independent of the absolute flux calibration. Moreover, as it is 
calculated within a wavelength region only $\approx 1000$\AA~ wide,
it will be relatively insensitive to both the
relative flux calibration and reddening.

\subsection{Definition of \RCa}
\label{sec:rca-1}

Following \cite{nugseq95}, we defined two zones in the
$3000$\AA--$4000$\AA\ region of the spectrum, centered on the \CaII\ 
$3650$\AA\ and $3933$\AA\ lines, as shown in the right hand side of
Fig.~\ref{fig:RCaDef}. \RCa\ is defined as the ratio of the $\approx
3933$\AA\ maximum, max$(F(3933))$, to the one at $\approx 3650$\AA,
max$(F(3650))$.
\begin{equation}
  \label{eq:48}
  \textrm{\RCa}=\frac{\mathrm{max}(F(3933))}{\mathrm{max}(F(3650))}
\end{equation}
The width of the zones searched for these maxima are displayed in 
Table~\ref{tab:rca}.
 
\begin{deluxetable}{cc}
\tablecolumns{2}
\tablewidth{0pc}
\tablecaption{\RCa\ Maximum Zones\label{tab:rca}}
\tablehead{
\colhead{Central Wavelength (\AA)}  & \colhead{Zone half-width (\AA)}}
\startdata
  3650 & 48 \\
  3933 & 65 \\
\enddata
\end{deluxetable}

The maxima are almost \emph{never} found at the exact wavelength of
the \CaII\ lines. This is expected since even if we had pure \CaII, 
effects other than resonant-scattering could move the feature's maximum
wavelength. Moreover, these features are line blends,
and not pure \CaII\ lines.

The study of these wavelength zones with \phoenix\ showed that the
dominant elements of these blends had lines displaying 
the same qualitative evolution as \RCa\ with luminosity. This leads us
to propose that the maxima ratio correlation with luminosity is only
a corollary of the existence of a correlation between the blends.
Hence, we defined an integral ratio we named \RCaS\ (the S
stands for surface)  as
\begin{equation}
  \label{eq:49}
  \textrm{\RCaS}=\frac{\int^{4012}_{3887}F_{\lambda}d\lambda}{\int^{3716}_{3620}F_{\lambda}d\lambda}
\end{equation}
The wavelength regions were empirically chosen in order to group all
the contributions with the same evolution with luminosity than
\RCa. We chose the wavelength limits in Eq.~\ref{eq:49} in order to
produce a spectral index that is well correlated to luminosity and 
robust to wavelength errors. 

Both the \RCa\ and \RCaS\ ratios, like \RSi,  are independent of the
absolute flux calibration, and not too sensitive to the relative
calibration quality or reddening.

\subsection{Dealing with Noise: }
\label{sec:dealing-with-noise}

The main merit of \RCaS\ over \RCa\ is the increased signal to
noise ratio. On the other hand, \RCa\ and \RSi\ will be
highly sensitive to noise since they rely on 
maxima and minima detections. In order to reduce this effect,
we fit each wavelength zone considered with a $4^{\textrm{th}}$
degree polynomial and search for their local maximum or minimum. 
 
Because we want \RSi\ and \RCa\ calculation algorithms to be
automatic and robust, we need to search wavelength regions wide
enough to be sure to find the right maxima and minima. But in these large
zones the spectral feature shapes are more complex than that of a second degree
polynomial. On the other hand, noise may generate oscillations 
which will be followed by a polynomial of too high a degree. We
chose $6^{\textrm{th}}$ degree polynomials as they 
empirically proved to be a good compromise between these two effects.

\section{Correlation of \RSi\ \& \RCa\  with Luminosity}
\label{sec:rsi--rca-1}

\begin{deluxetable}{llc}
\tablecolumns{3}
\tablewidth{0pc}
\tablecaption{List of Observed \supernovae   \label{tab:snused}}
\tablehead{
\colhead{SN}    &   \colhead{Epoch (day wrt max)}   &
\colhead{M$_B$} }
\startdata
  SN 1998aq &  -3, 0, 1, 2, 3, 4 & -19.56 \\
  SN 1981B  &    0$^*$ &  -19.54   \\
  SN 1998bu & -4, -2, -1 & -19.49 \\
  SN 1994D  &   2$^*$, 3$^*$, 4$^*$, 5$^*$, -3, -4, -5  & -19.47\\
  SN 1996X  &   -4, -2, 4, 0$^*$, 1$^*$ & -19.43 \\            
  SN 1989B  &   -1, -5 &  -19.42\\
  SN 1992A  &   0, -1, -5, 3, 5  & -19.32\\
  SN 1986G  &   -1, -3, -5$^*$, 1, 3 &  -19.16 or -17.58 \\
\enddata
\tablenotetext{a}{\Supernovae used in our simulation. Absolute magnitudes
are taken from from \citet{reindl05}.  The spectra from the
epochs denoted with an asterisk lacked coverage in the blue and were
excluded from the calculations of \RCa\ and \RCaS\ (see text).}
\end{deluxetable}

We applied these definitions to the public \supernovae spectra we
were able to gather, listed in
Table~\ref{tab:snused}. The absolute magnitudes of \citet{reindl05}
are derived for a common arbitrary value of the Hubble Constant
$H_0=60$\kmps~Mpc$^{-1}$. The values of the reddening and $\delta
m_{15}$ were chosen to minimize deviations from the Hubble law for a
set of 66 \SNeIa in the Hubble flow. While other prescriptions for
determining $M_B$ would give somewhat different results, with our
small sample it is advantageous to use a single consistent
prescription to determine $M_B$ rather than relying on a variety of
methods to determine the distances to nearby \SNeIa. When present
surveys that will construct large homogeneous samples of \SNeIa in the
Hubble flow are complete, it will be simple and necessary to redo our
analysis. Nevertheless, we don't  expect to alter our qualitative
conclusions.  \citet{reindl05} were unable to determine a unique
solution for the reddening to SN~1986G, and thus we have  two
$M_{B}$ values for SN 1986G which we carry through in our analysis.
In the
following, we will calculate results for both values of the absolute
$B$ magnitude of SN~1986G, we will refer to the $M_{B}=-17.58$ case as 
``Case A'',
and $-19.16$ as ``Case B''. Since the analysis of
\citet{reindl05} is the most recent and careful, we
consider Case B to be the more likely value for $M_{B}$ and we will
present plots of our results for Case B only, however since the
reddening correction is uncertain and we are dealing with only a small
number of supernovae, we will present our results for both cases in
tabular form. For our calibration, we restrict ourselves to ``Branch
Normals'' which we define to be \SNeIa with $-19.0 < M_B < -19.8$,
thus in Case A, SN~1986G \emph{is not} included in the set of
calibrators, whereas in Case B it is.
As more spectra of SNe~Ia
with well-measured values of $M_{B}$ become available they are easily
added to our analysis.

Even though \RSi\ and \RCa\ were originally defined for \SNeIa spectra
at maximum light, we used spectra ranging from $5$~days prior to
maximum to $5$~days after maximum.  We examined various epochs around
maximum in order to determine whether or not the date of the spectra
has a critical impact on the ratio's correlation with
luminosity. Moreover, since space programs like \snap\ will
only take one spectrum per supernova, we wanted to quantify the time
window where spectra are usable for luminosity measures with line
ratios. Note that JEDI proposes to take multiple spectra of each SN,
at 5-day intervals in the observer's frame.

On the other hand, because of time dilatation, the rise time for
\supernovae at $z \approx 1$ is twice  as slow in the observer's
rest frame. Relaxing the time constraint to $\pm 5$ days around
maximum was thus conservative.

\subsection{\RCa\ \& \RCaS\ Correlation with Luminosity}
\label{sec:rca-correlation-with}

\begin{figure}
\centering
  \includegraphics[width = 0.8\textwidth]{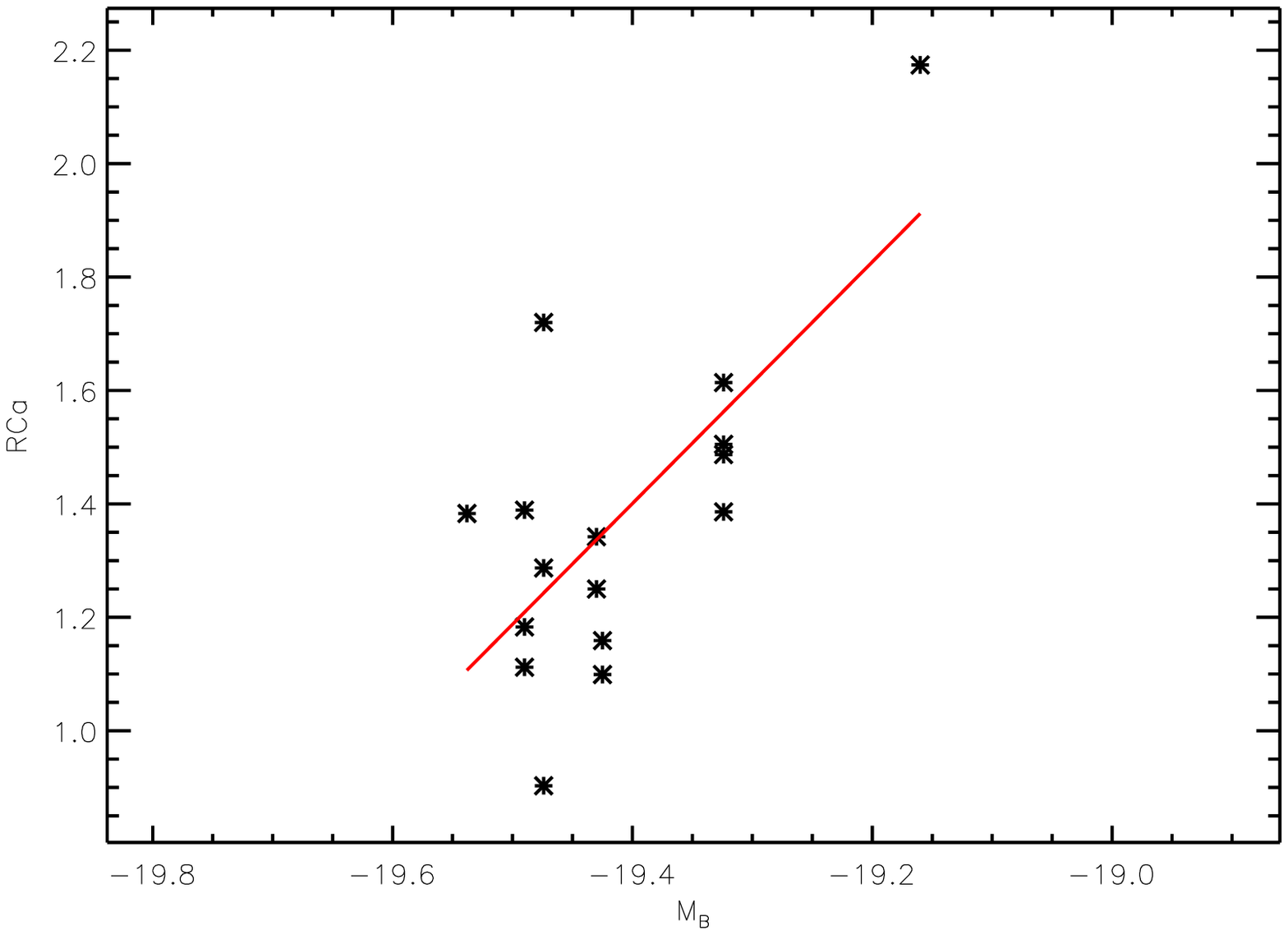}
  \caption[\RCa\ correlation with luminosity]{\RCa
      calculated using \supernovae listed in
      Table~\ref{tab:snused} (Case B). The line is the result of the
      linear regression.}
  \label{fig:RCareal}
\end{figure}

\begin{figure}  
\centering
  \includegraphics[width = 0.8\textwidth]{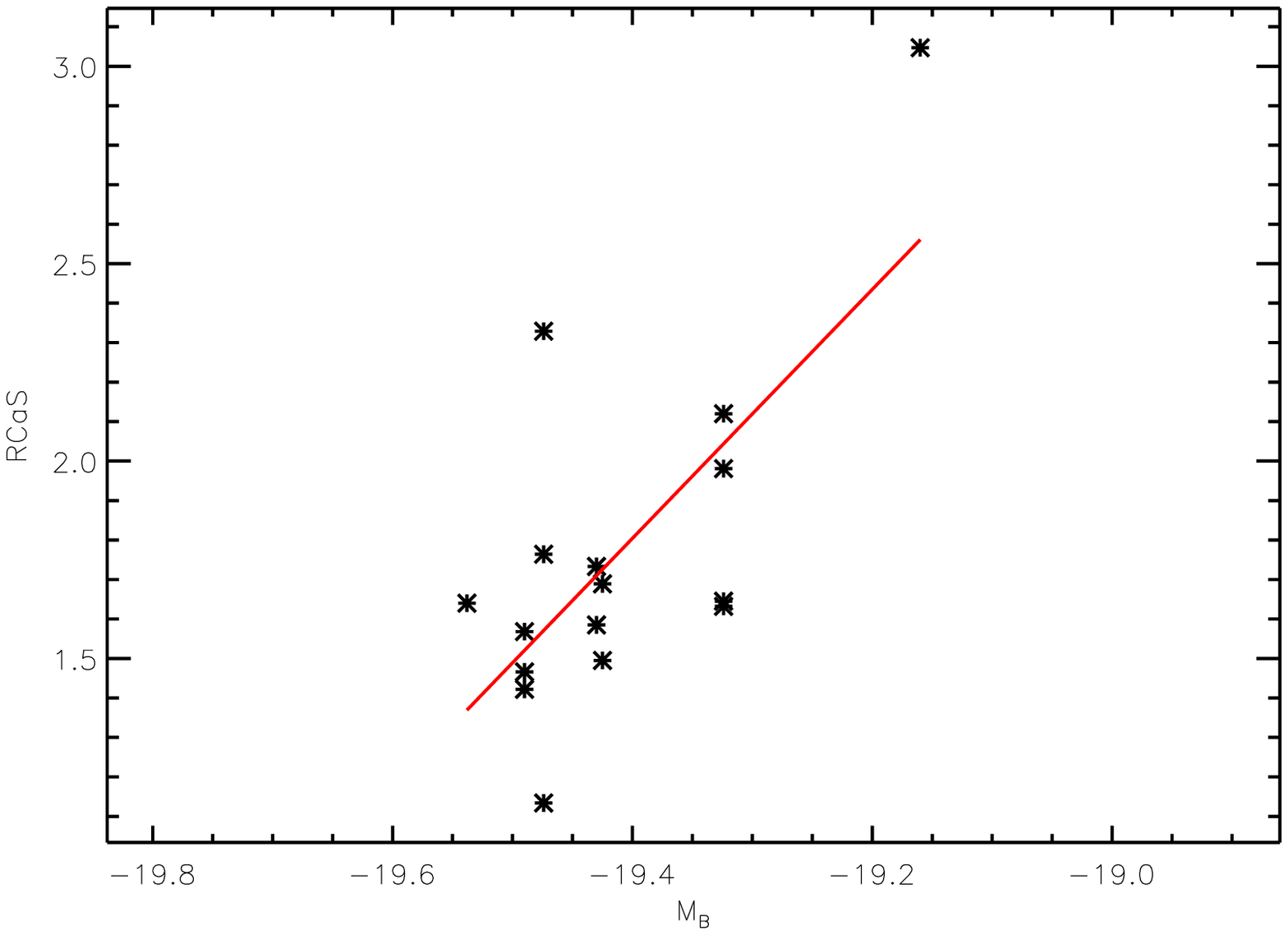}
  \caption[\RCaS\ correlation with luminosity]{\RCaS
      calculated using \supernovae listed in
      Table~\ref{tab:snused} (Case B). The line is the result of the
      linear regression.}
  \label{fig:RCaSreal}
\end{figure}

Some of the spectra listed in Table~\ref{tab:snused} lacked the
wavelength coverage needed to calculate \RCa\ and \RCaS, those epochs
are indicated with an asterisk. Neglecting these, the results for Case B
are plotted in Figs.~\ref{fig:RCareal} and \ref{fig:RCaSreal}.

The \RCa\ values for maximum light spectra agree with \cite{nugseq95} to
within a few percent. In order to calibrate the luminosity relations,
we computed the linear regression of \RCa\ and \RCaS\ values with
respect to blue magnitude. This linear regression should be valid in
the restricted set of ``Branch Normals''. 3-D deflagration models have
difficulty reproducing the very low nickel mass of SN~1991bg-like
\supernovae and we illustrate below that synthetic spectra of W7
\citep{nomw7} indicate that there may be a departure of the spectral
ratios from linearity
for 1991T-like \supernovae. Thus, by restricting our calibrators to Branch
Normals we \emph{might} be using a more similar population. From the
linear regression we obtain 
\begin{eqnarray}
  \label{eq:28}
    \textrm{\RCa} & = a_{\textrm{\RCa}}~  M_{B} + b_{\textrm{\RCa}}  \\
    \textrm{\RCaS} & = a_{\textrm{\RCaS}}~  M_{B} + b_{\textrm{\RCaS}}  
\end{eqnarray}

We list the resulting parameters in
Table~\ref{tab:RCaRCaSLinReg2}. Since 
we have no information on the experimental 
errors, we set them arbitrarily equal to unity.

We also calculated the standard deviation $\sigma_{\textrm{\RCa}}$ for
\RCa\ and \RCaS, defined as the 
quadratic mean of the distance of each supernova from the linear
regression, and have listed them in the same table. This definition of
the error combines measurement errors, time dependence, and intrinsic
dispersion. 

\begin{deluxetable}{lccc}
\tablecolumns{4}
\tablewidth{0pc}
\tablecaption{\RCa\ \& \RCaS\ luminosity measure
  precision\label{tab:RCaRCaSLinReg2}} 
\tablehead{
& \colhead{$\sigma_{\textrm{\RCa}}$}    &   \colhead{$a_{\textrm{\RCa}}$}   &
\colhead{$\sigma_{M_{B}}$}}
\startdata
\RCa\ (Case B)   &      0.20      &    2.1   &      0.10 \\
\RCaS\ (Case B)  &      0.31      &    3.1  &      0.10 \\
\RCa\ (Case A)  &    0.19     &  1.2    &      0.16 \\
\RCaS\ (Case A) &    0.39     &  1.4    &      0.28 \\
\enddata
\end{deluxetable}

\paragraph{Luminosity measure precision using \RCa\ or \RCaS: }
\label{sec:lumin-meas-prec}

With the linear regression calibration of \RCa\ and \RCaS\ with respect
to blue magnitude and the associated variance, we where able to
estimate the luminosity precision measure using these ratios as
follows: 
\begin{eqnarray*}
\label{eq:lumerr}
  \textrm{\RCa} & = a_{\textrm{\RCa}}  M_{B} + b_{\textrm{\RCa}}  \\
  \sigma_{\textrm{\RCa}} & = a_{\textrm{\RCa}}  \sigma_{M_{B}} \\
   \sigma_{M_{B}} & = \frac{\sigma_{\textrm{\RCa}}}{a_{\textrm{\RCa}}}\\
  \frac{\Delta L}{L}&  \approx \frac{\sigma_{\textrm{\RCa}}}{a_{\textrm{\RCa}}}
\end{eqnarray*}

We display in Table~\ref{tab:RCaRCaSLinReg2} the present measurement accuracy on
the blue magnitude as estimated for \RCa\ and \RCaS. We find in Case B
that the accuracy of the method reaches
the $0.1$ magnitude level.

\subsubsection{\RSi\ correlation with luminosity: }
\label{sec:rsi-correlation-with}

\paragraph{The correlation calibration}
\label{sec:corr-calibr}

\begin{figure}
\centering
  \includegraphics[width = 0.8\textwidth]{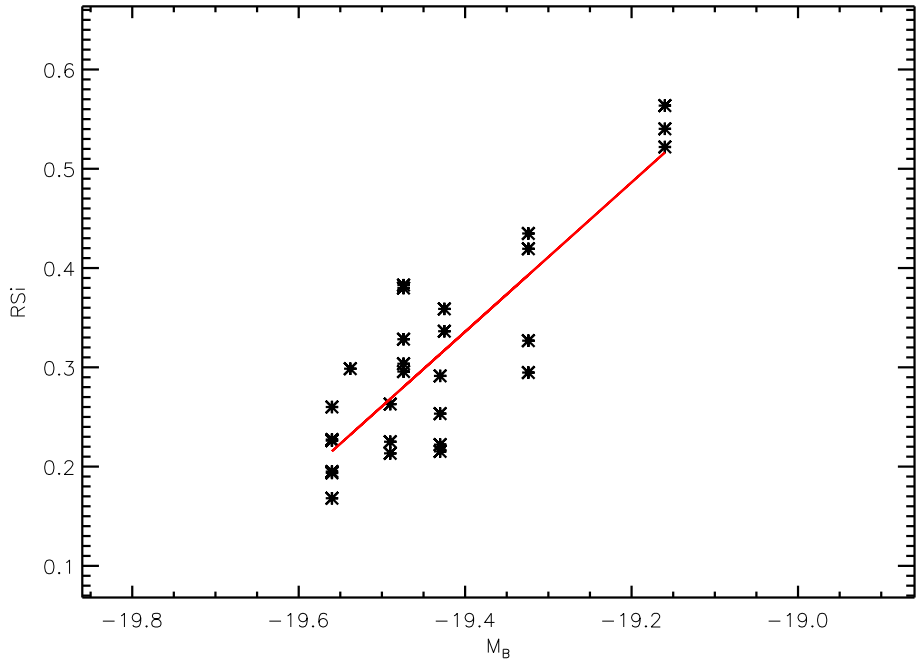}
  \caption[\RSi\ correlation with luminosity]{\RSi
      calculated using \supernovae
    in Table~\ref{tab:snused} (Case B). The line is the result of our
    linear regression.
}
  \label{fig:RSireal}
\end{figure}

Fig.~\ref{fig:RSireal} shows the \RSi\ values calculated with
\supernovae from Table~\ref{tab:snused}, including spectra at all
epochs. We excluded a negative 
\RSi\ value from the
$-5$ day SN 1994D spectrum, for which the blue trough associated 
with the \SiIIblue\ feature does not exist.  
The \RSi\ ratio is of course mathematically well defined even in this
peculiar case, but since this point falls out of the general trend and
increases the \RSi\ slope, we removed it from the linear
regression. This selection is compatible with observational
constraints, as such spectra are easily identified when the signal to
noise is large enough to measure \RSi. 

The linear regression and the corresponding relative dispersion for
this selection of \supernovae is also shown in Fig.~\ref{fig:RSireal}.
\begin{equation}
  \label{eq:29}
  \textrm{\RSi}= a_{\textrm{\RSi}} M_{B} + b_{\textrm{\RSi}}
\end{equation}

We also calculated the associated standard deviation and
found it to be $\sigma_{\textrm{\RSi}} \approx 0.09$
(Case B) or $\sigma_{\textrm{\RSi}} \approx 0.06$ (Case A). 
This variance includes the measurement errors, time dispersion, as
well as the intrinsic 
dispersion of the \supernovae.

\paragraph{Luminosity measure precision: }
\label{sec:lumin-meas-prec-1}

As in \S\ref{sec:lumin-meas-prec} we estimate the accuracy of the
luminosity determination using \RSi\ directly from the dispersion with 
respect to the linear regression. Table \ref{tab:RSiaccuracy}
summarizes the results. 

\begin{deluxetable}{clcc}
\tablecolumns{4}
\tablewidth{0pc}
\tablecaption{\RSi\ luminosity measure precision\label{tab:RSiaccuracy}}
\tablehead{
& \colhead{$\sigma_{\textrm{\RSi}}$}    &   \colhead{$a_{\textrm{\RSi}}$}   &
\colhead{$\sigma_{M_B}$}}
\startdata
\RSi\ (Case B)  & 0.05 & 0.75 &     0.07\\
\RSi\ (Case A) &      0.06 &     0.59 &      0.10 \\
\enddata
\end{deluxetable}

\section{\RSiSu: A New Spectral Ratio}
\label{sec:rsisu-new-spectral}

\begin{figure}
  \includegraphics[width = 0.8\textwidth, clip]{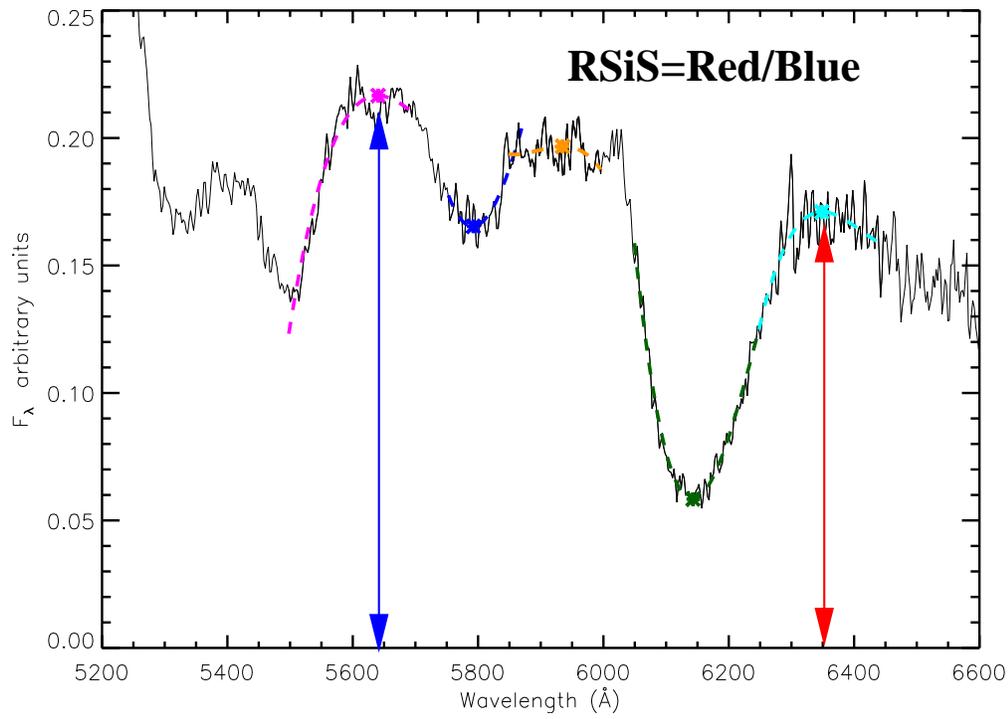}
  \caption[\RSiSu]{The definition of the method used for determining
    \RSiSu. The arrows indicate the maxima
     used to compute \RSiSu.}
  \label{fig:RSiSuDef}
\end{figure}

Due to the \RSi\ correlation with luminosity, we have focused on the
formation of features in the \RSi\ zone.
More will be said about
the line formation in this spectral region in a forthcoming
publication (S.~Bongard et al., in preparation), but the
clarification of the spectrum formation 
process led us to devise a new line ratio indicator which we
call \RSiSu. 

\RSiSu\ is defined as the ratio of two maxima. The first maximum is the
one used in the \RSi\ calculation, located around {\SiIIred,} and the
second one is the bluer peak of due to sulfur (hence the name Si+S a
mixture of silicon and sulfur) located in the $5600-5700$\AA{}
region as shown in Figure~\ref{fig:RSiSuDef}. The zones where we look
for these maxima are listed in Table~\ref{tab:rsisu}. As was the case
for \RSi\ and \RCa, we used a \xth{4} order polynomial to fit the
regions searched for the maxima, in order to increase 
robustness of the ratio with respect to Poisson noise.

\begin{deluxetable}{lcc}
\tablecolumns{3}
\tablewidth{0pc}
\tablecaption{\RSiSu\ zones\label{tab:rsisu}}
\tablehead{
& \colhead{Lower Wavelength (\AA)}    &   \colhead{Upper Wavelength (\AA)}}
\startdata
  First maximum&6200 & 6450 \\
  Third maximum&5500 & 5700 \\
\enddata
\end{deluxetable}

We also define the integral ratio \RSiSuS\ as:
\begin{equation}
  \label{eq:51}
  \mbox{\RSiSuS}=\frac{\int_{5500}^{5700}F_{\lambda}d\lambda}{\int_{6450}^{6200}F_{\lambda}d\lambda}.    
\end{equation}

Once again, the simplicity of this integral ratio calculation and its
insensitivity to noise makes it an interesting alternative to the line
fitting method.

\section{\RSiSu\ correlation with luminosity}
\label{sec:rsisu-corr-with}

\begin{figure}
\centering
  \includegraphics[width = 0.8\textwidth]{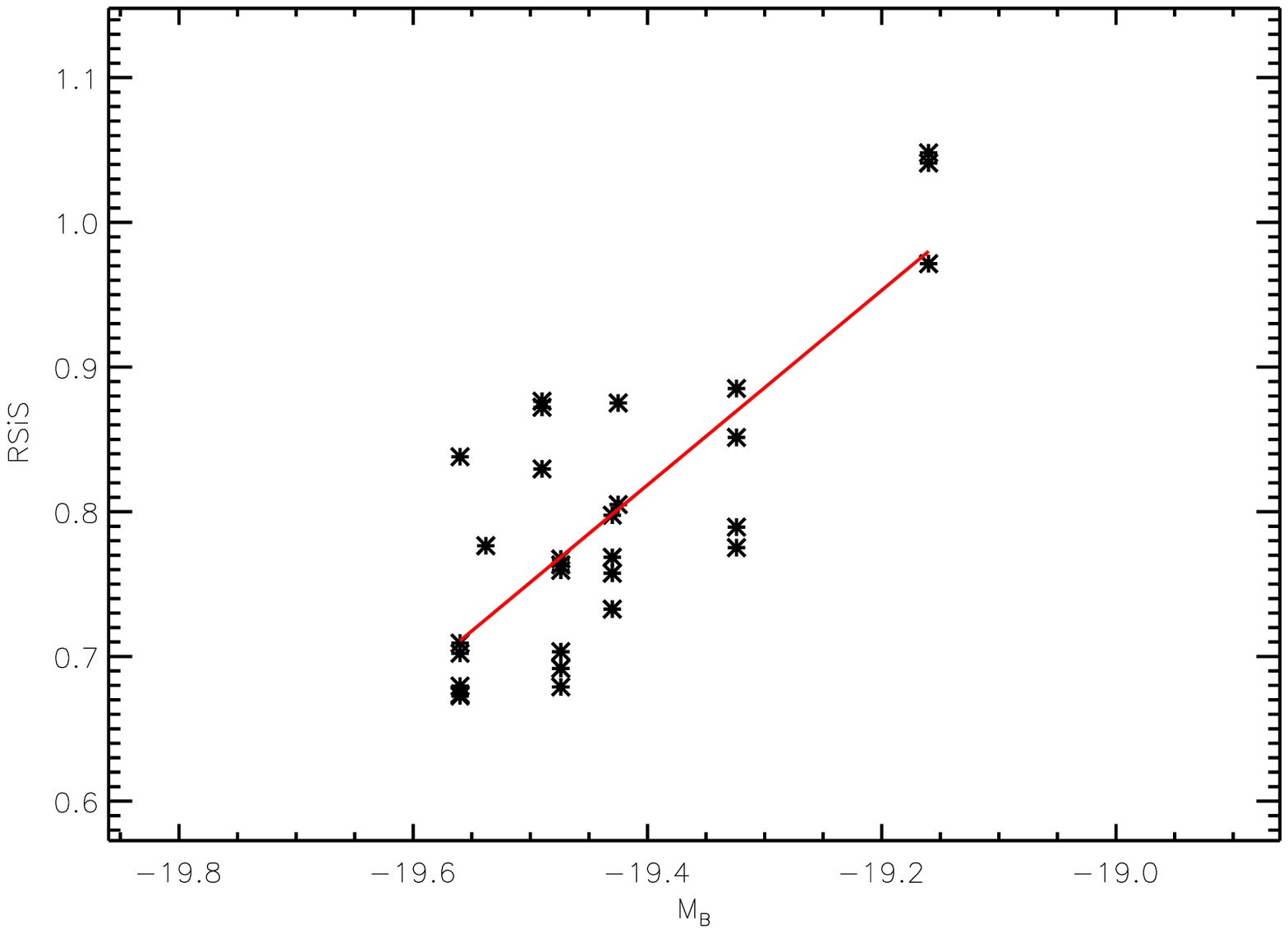}
  \caption[\RSiSu\ correlation with luminosity, no 91bg]{\RSiSu\ (Case B)
    calculated using the \supernovae of 
    Table~\ref{tab:snused}. The line is the result of our linear regression.}
  \label{fig:RSiSureal}
\end{figure}

\begin{figure}
\centering
  \includegraphics[width = 0.8\textwidth]{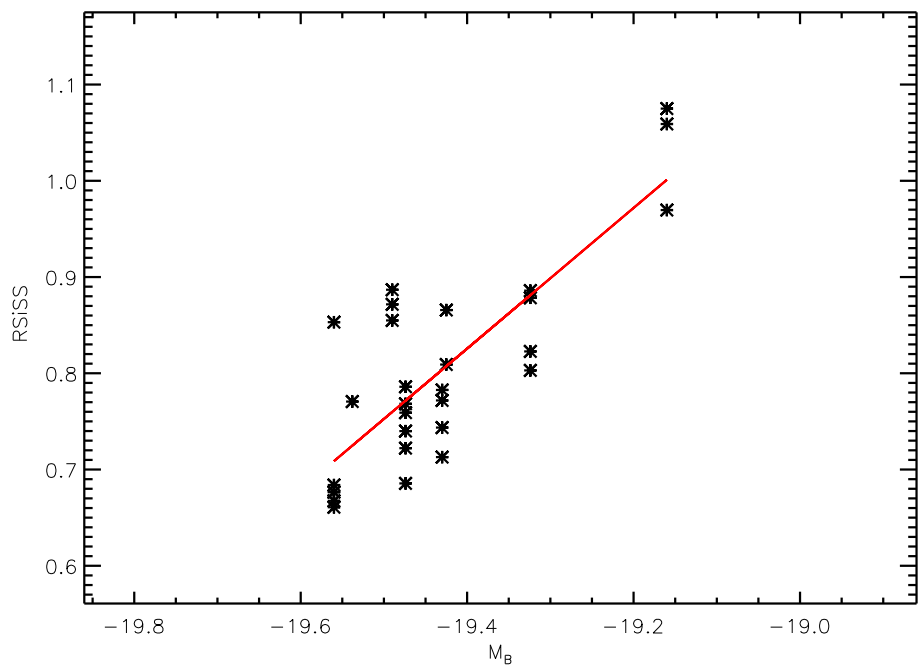}
  \caption[\RSiSuS\ correlation with luminosity, no 91bg]{\RSiSuS\ (Case B)
    calculated using the \supernovae of 
    Table~\ref{tab:snused}. The line is the result of our linear regression.}
  \label{fig:RSiSuSreal}
\end{figure}

We plot the linear regression of \RSiSu\ correlation with luminosity in
Fig.~\ref{fig:RSiSureal} 
We also plot in
Fig.~\ref{fig:RSiSuSreal} the  \RSiSuS\ 
correlation with luminosity. Table \ref{tab:RSiSuPrec}
summarizes the results for both quantities.

\subsection{Luminosity measure precision}
\label{sec:lumin-meas-prec-2}

We summarize the different slopes and
dispersion values for \RSiSu\ and \RSiSuS\ calculated with or without
the time dispersion correction in Tables~\ref{tab:RSiSuPrec} and
\ref{tab:RSiSutime}.

\begin{deluxetable}{llcc}
\tablecolumns{4}
\tablewidth{0pc}
\tablecaption{\RSiSu\ luminosity measure precision\label{tab:RSiSuPrec}}
\tablehead{
& \colhead{$\sigma$}    &
\colhead{$a$}   &
\colhead{$\sigma_{M_{B}}$}}
\startdata
\RSiSu\ (Case B)     & 0.06    &  0.67  &  0.09  \\
\RSiSuS\ (Case B)   & 0.06    &  0.73  &  0.09  \\
\RSiSu\  (Case A)    & 0.06   &  0.43   &  0.14 \\
\RSiSuS\  (Case A)   & 0.07   &  0.53   &  0.11 \\
\enddata
\end{deluxetable}


With this spectral indicator family, even without time correction, we
have augmented the blue magnitude accuracy by a factor of two compared
to \RCa. In Case A, \RSi\ and
\RSiSu\ have similar accuracies; however, in Case B
\RSiSu\ is slightly better.
Since \RSiSu\ and \RSi\ cannot be considered as 
independent, we still have only two independent spectral
indicators with \RCa. But since one of them is as accurate as the
light curve method, coupled with enough \supernovae they have the
possibility to constrain evolutionary effects. 

\section{Time Evolution of Spectral Indices}
\label{sec:time-evolution}

\begin{figure}
\centering
  \includegraphics[width = 0.8\textwidth]{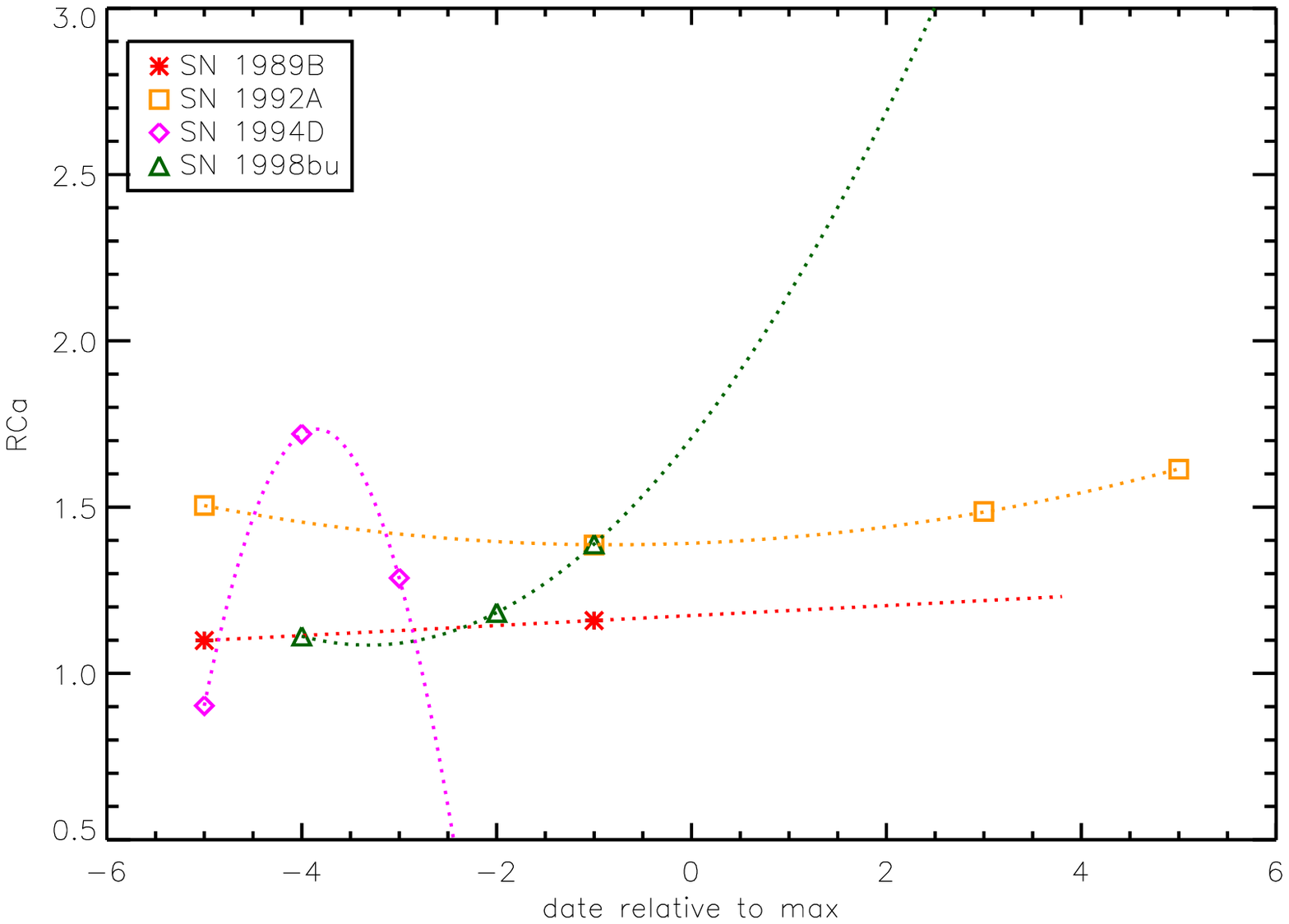}
  \caption[\RCa\ time evolution]{\RCa
    time evolution} 
  \label{fig:RCatime}
\end{figure}

\begin{figure}
\centering
  \includegraphics[width = 0.8\textwidth]{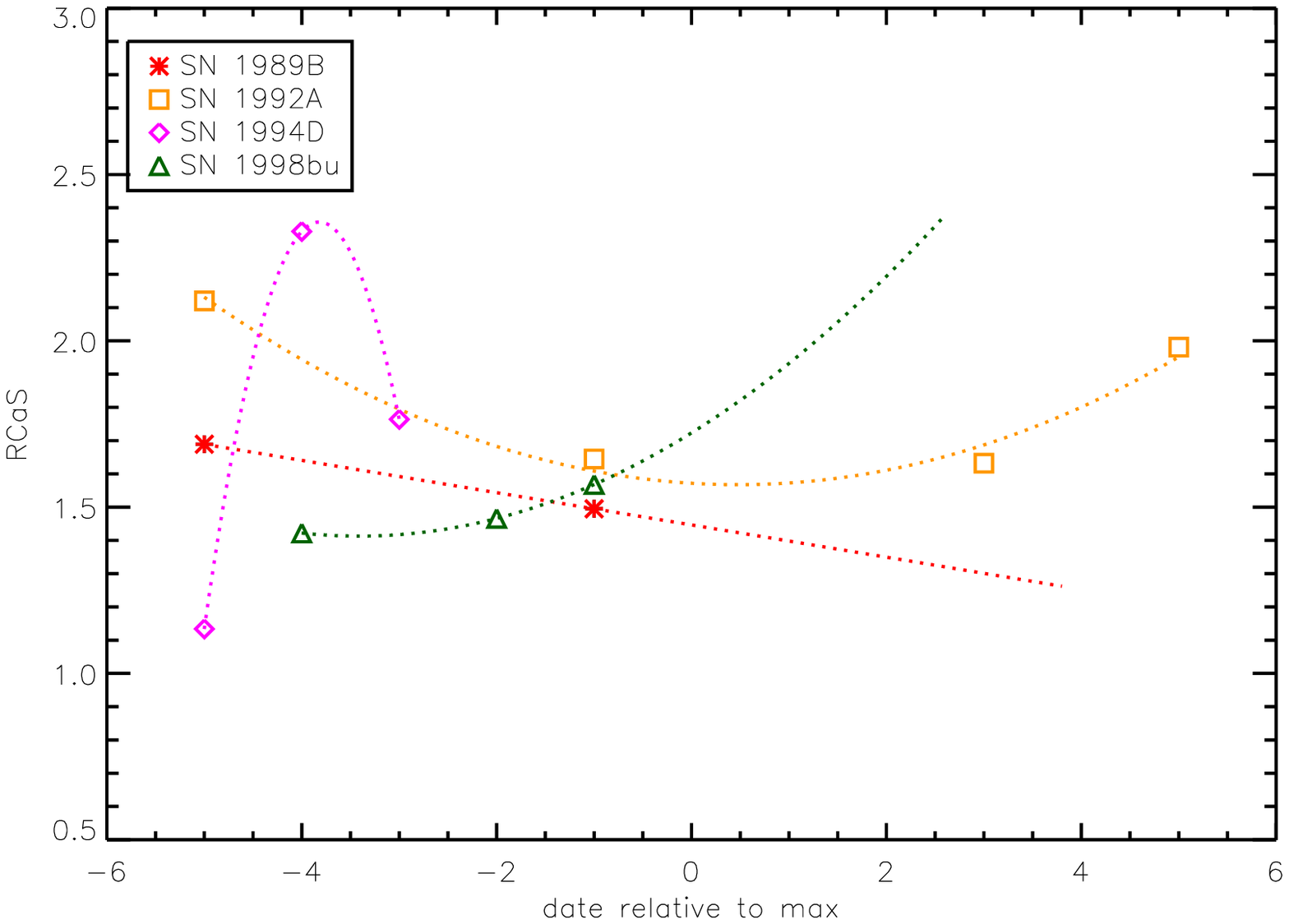}
  \caption[\RCaS\ time evolution]{
    \RCaS\ time evolution} 
  \label{fig:RCaStime}
\end{figure}

We plot in Fig.~\ref{fig:RCatime} and Fig.~\ref{fig:RCaStime} the
\RCa\ and \RCaS\ time 
dependence for each supernova. Whenever enough data points were available, we
interpolated the $t=0$ point using a $2^{\textrm{nd}}$ degree
polynomial. For the cases with two points only, we used a straight line.

The supernova SN 1994D had enough points for a \snd\ order polynomial
interpolation, but since all of them are for $t \leq -3$ days, and
since the trend is different from the other \supernovae, we considered
the $t=0$ point to be too much of an extrapolation to be used.
Whether the SN 1994D dispersion in \RCa\ or \RCaS\  is intrinsic or due to
measurement uncertainties is still an open question.

Replacing each \RCa\ and \RCaS\ by the corresponding $t=0$ calculated
value, we recomputed the linear regression, and the
associated standard deviation.
These results are shown in Table~\ref{tab:RCaRCaSLinRegTime}. 

\begin{deluxetable}{lccc}
\tablecolumns{4}
\tablewidth{0pc}
\tablecaption{\RCa\ and \RCaS\ interpolated linear
  regression\label{tab:RCaRCaSLinRegTime}}
\tablehead{
& \colhead{$\sigma$}  &
\colhead{$a$} & \colhead{$\sigma_{M_B}$}}
\startdata
\RCa\ (Case B) &   0.26  &   0.93 &     0.27 \\
\RCaS\ (Case B) &  0.35  &   1.19 &     0.30\\
\RCa\ (Case A) &   0.23  &   0.48 &     0.48 \\
\RCaS\ (Case A) &  0.24  &   0.75 &     0.31 \\
\enddata
\tablenotetext{a}{\RCa\ and \RCaS\ linear regression on the observed
  supernova spectra with sufficient wavelength coverage using the
  values interpolated to $t=0$.}
\end{deluxetable}

More spectra are needed in order to show if there is a correlation in
the evolution \RCa\ or \RCaS\ with epoch, but it is apparent that there
is no uniform trend with time, and the attempted correction to maximum
light has increased the dispersion from Table~\ref{tab:RCaRCaSLinReg2}
to Table~\ref{tab:RCaRCaSLinRegTime}.  Therefore, it is conservative
to treat 
the time dependence of \RCa\ and \RCaS\ as an uncontrolled dispersion.

\begin{figure}
\centering
  \includegraphics[width = 0.8\textwidth]{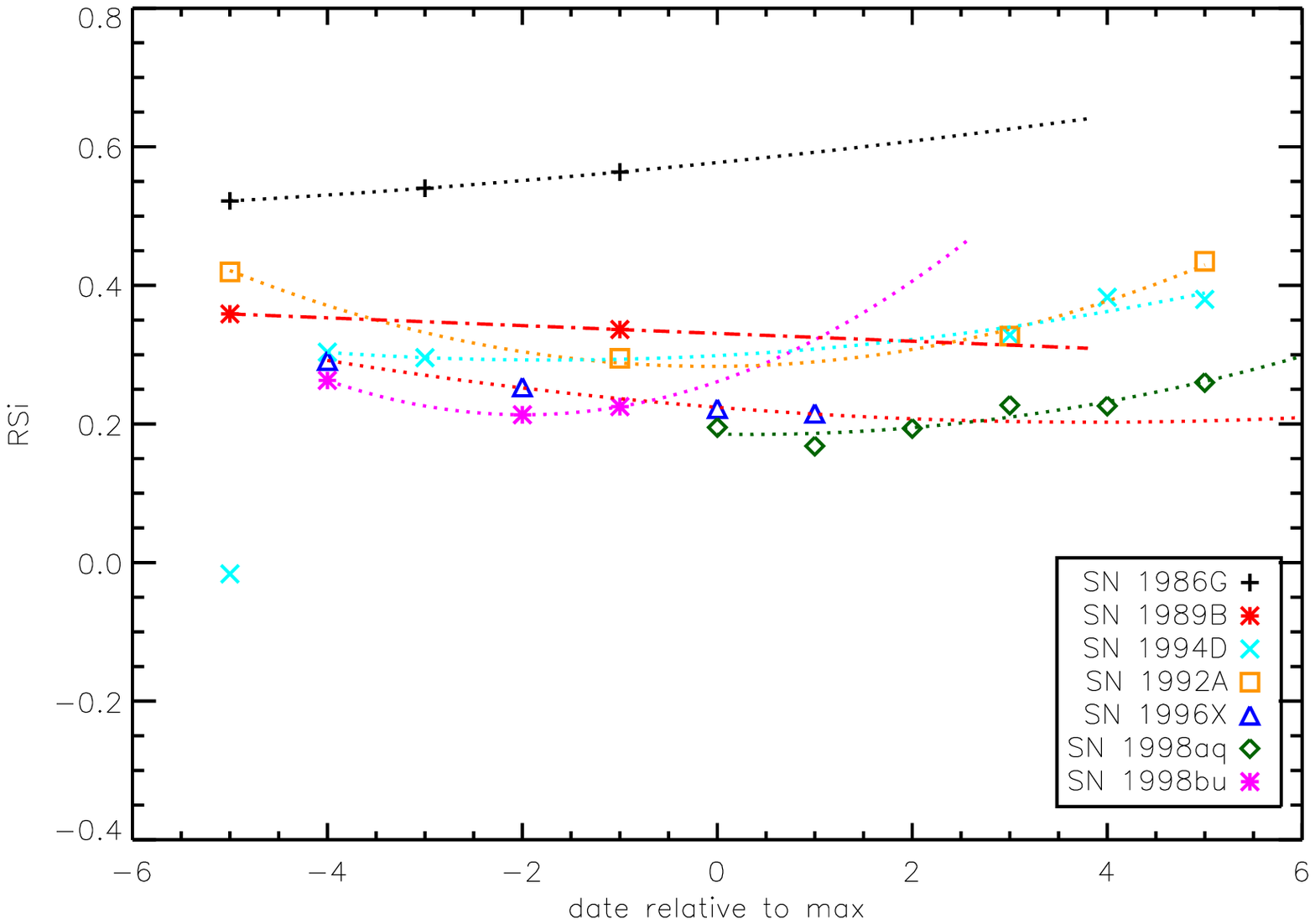}
  \includegraphics[width = 0.8\textwidth]{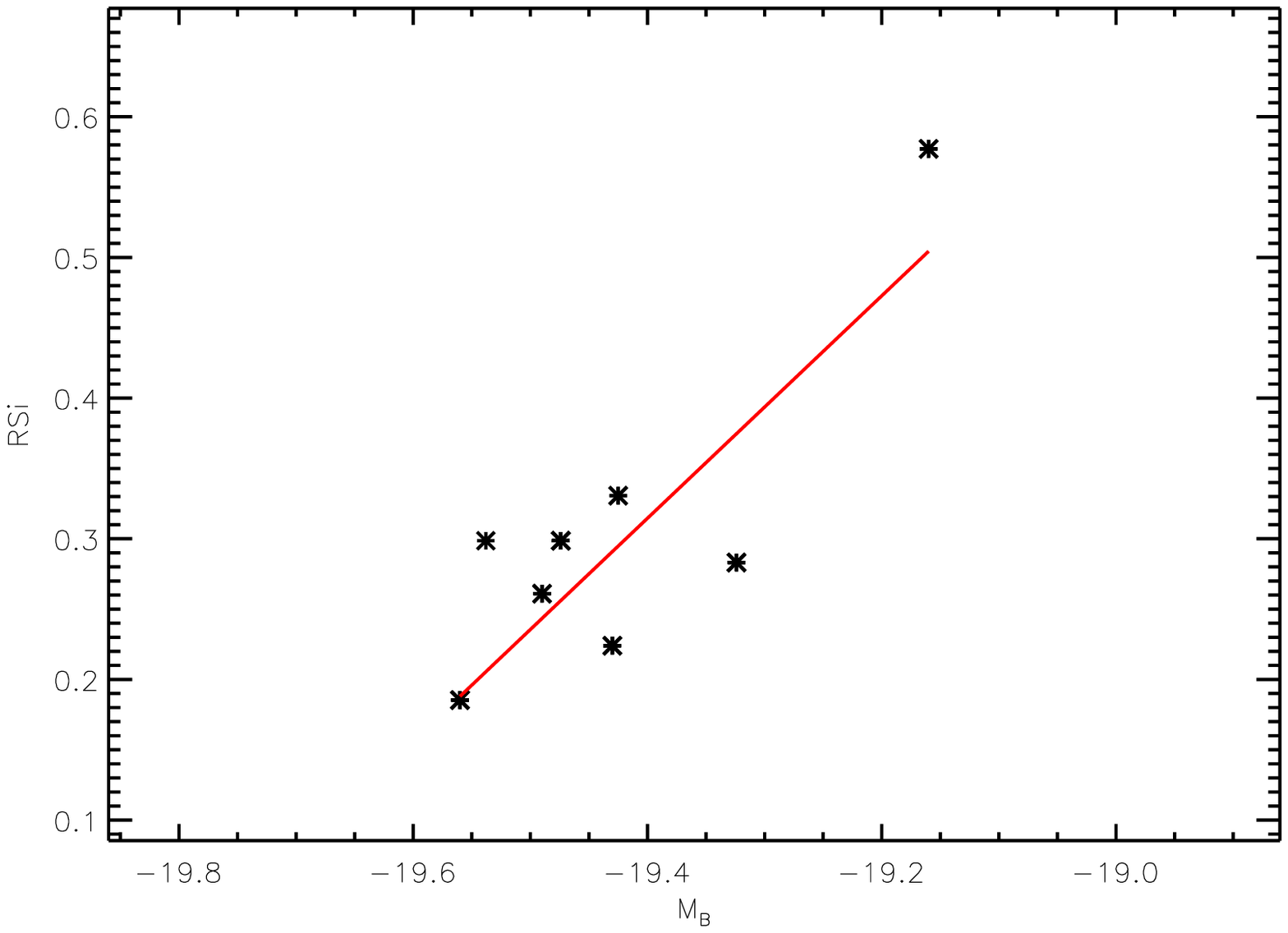}
  \caption[\RSi\ time dependence]{\RSi\ time 
    evolution (top panel, Case B) and \RSi\ correlation with
    values interpolated to $t=0$ (bottom panel).}
  \label{fig:RSitime}
\end{figure}

Fig.~\ref{fig:RSitime} displays the \RSi\ time dependence of the
\supernovae we used. We calculated a quadratic regression
for each supernova with three or more \RSi\ values. We then
replaced \emph{each} \RSi\ value by the corresponding value at $t=0$
obtained via either interpolation or extrapolation. The new linear regression
calculated for \RSi\ correlation with luminosity is also plotted in
Fig.~\ref{fig:RSitime}. We calculated the same correction for the less
favored luminosity choice for SN 1986G (Case A).

\begin{deluxetable}{clcc}
\tablecolumns{4}
\tablewidth{0pc}
\tablecaption{\RSi\ interpolated luminosity measure precision\label{tab:RSiaccuracytzero}}
\tablehead{
& \colhead{$\sigma_{\textrm{\RSi}}$}    &   \colhead{$a_{\textrm{\RSi}}$}   &
\colhead{$\sigma_{M_B}$}}
\startdata
\RSi\ (Case B) & 0.06 & 0.79 &     0.07\\
\RSi\ (Case A)  &      0.04 &     0.35 &      0.12 \\
\enddata
\tablenotetext{a}{\RSi\ linear regression on the observed
  supernova spectra with sufficient wavelength coverage using the
  values interpolated to $t=0$.}
\end{deluxetable}

For the favored SN 1986G luminosity case (Case B)
the extrapolation to $t=0$
 significantly improves the measurement, as seen in
 Table~\ref{tab:RSiaccuracytzero} and lowers the 
dispersion to 
$0.1$ blue magnitude so that the 
\RSi\ ratio probes  ``Branch normal'' \SNeIa
luminosities with good accuracy.  

The standard deviation calculated with respect to this new regression, still
removing the events with negative \RSi, is now $\sigma_{\textrm{\RSi}}
\approx 0.07$ (Case B) or $\sigma_{\textrm{\RSi}} \approx 0.12$ (Case A).
The top panel of Fig.~\ref{fig:RSitime}  shows that there is no
general time dependence trend for our \supernovae sample.

\begin{figure}
\centering
  \includegraphics[width = 0.8\textwidth]{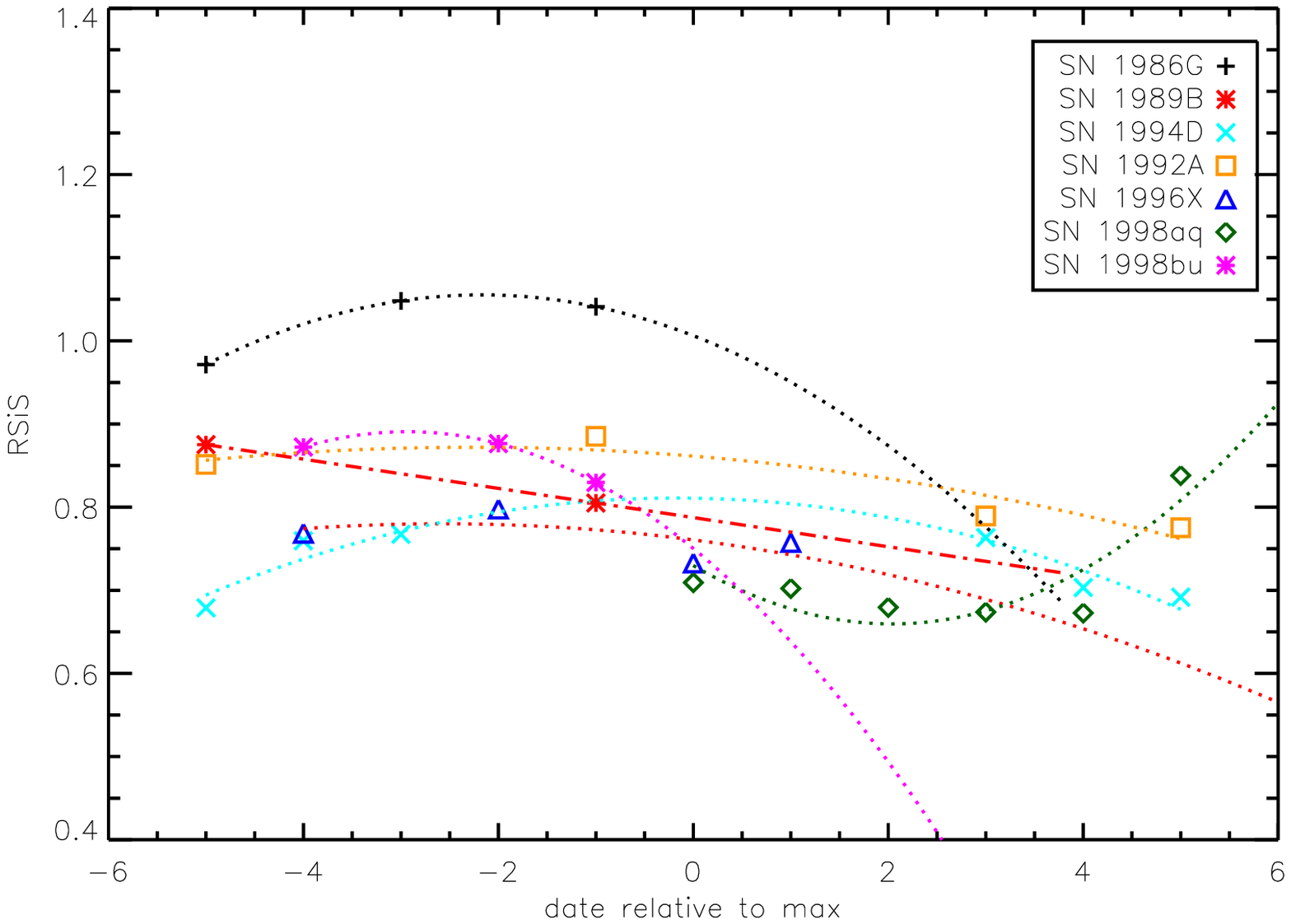}
  \includegraphics[width = 0.8\textwidth]{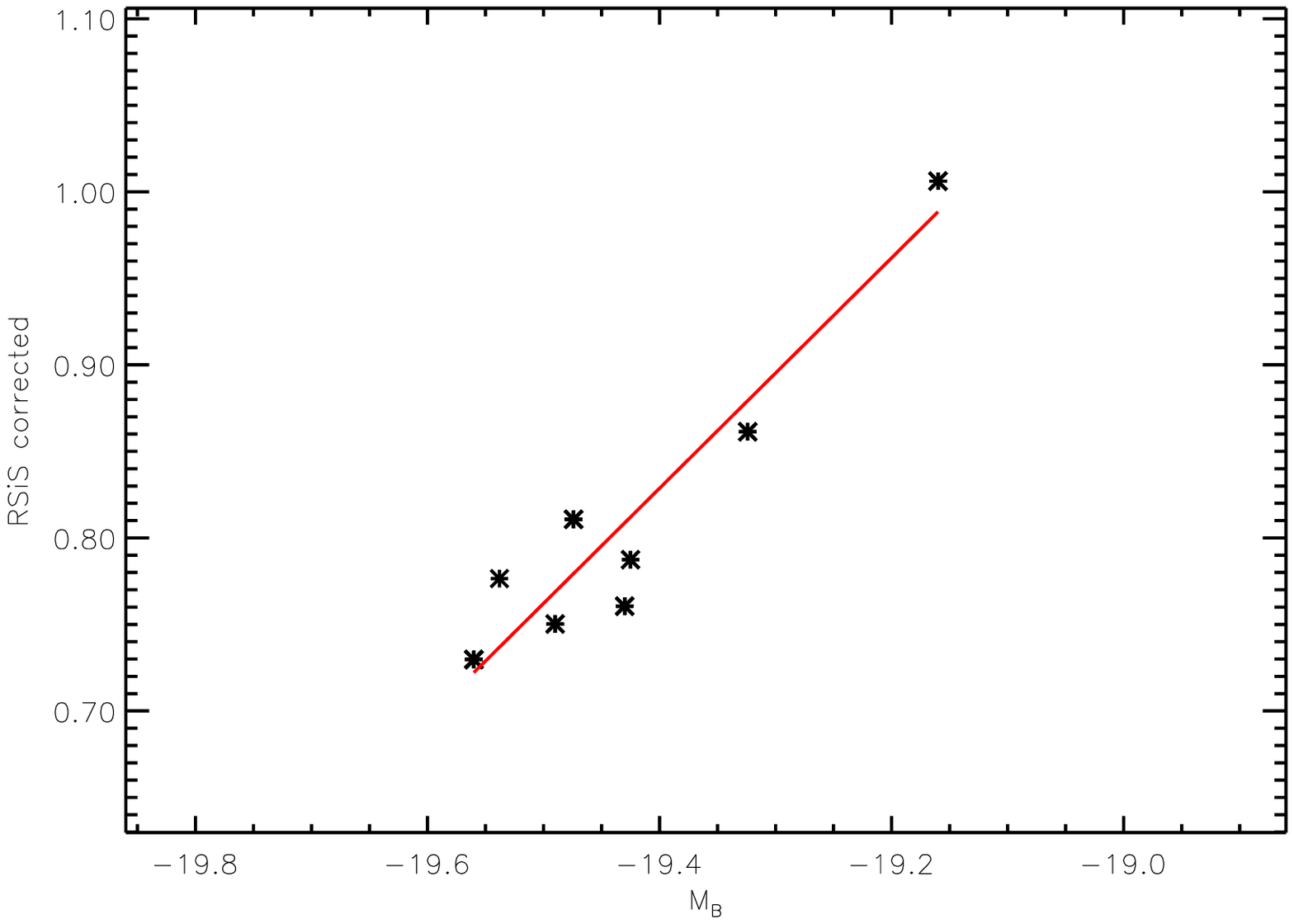}
  \caption[\RSiSu\ time evolution]{
    \RSiSu\ time evolution (top panel), \RSiSu\  corrected for the
    time dependence (bottom panel).}
  \label{fig:RSiSurealTime}
\end{figure}

\begin{figure}
\centering
  \includegraphics[width = 0.8\textwidth]{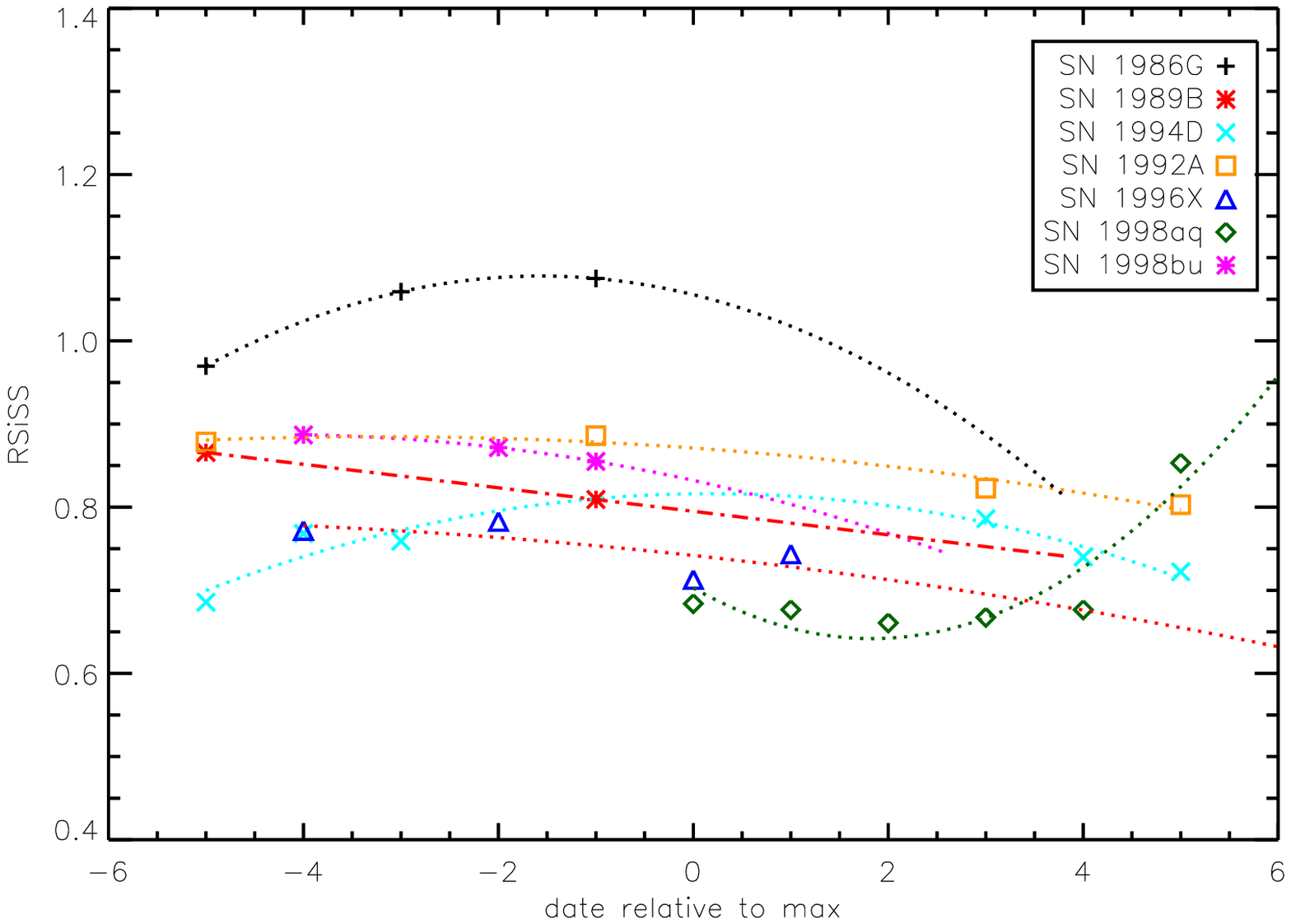}
  \includegraphics[width = 0.8\textwidth]{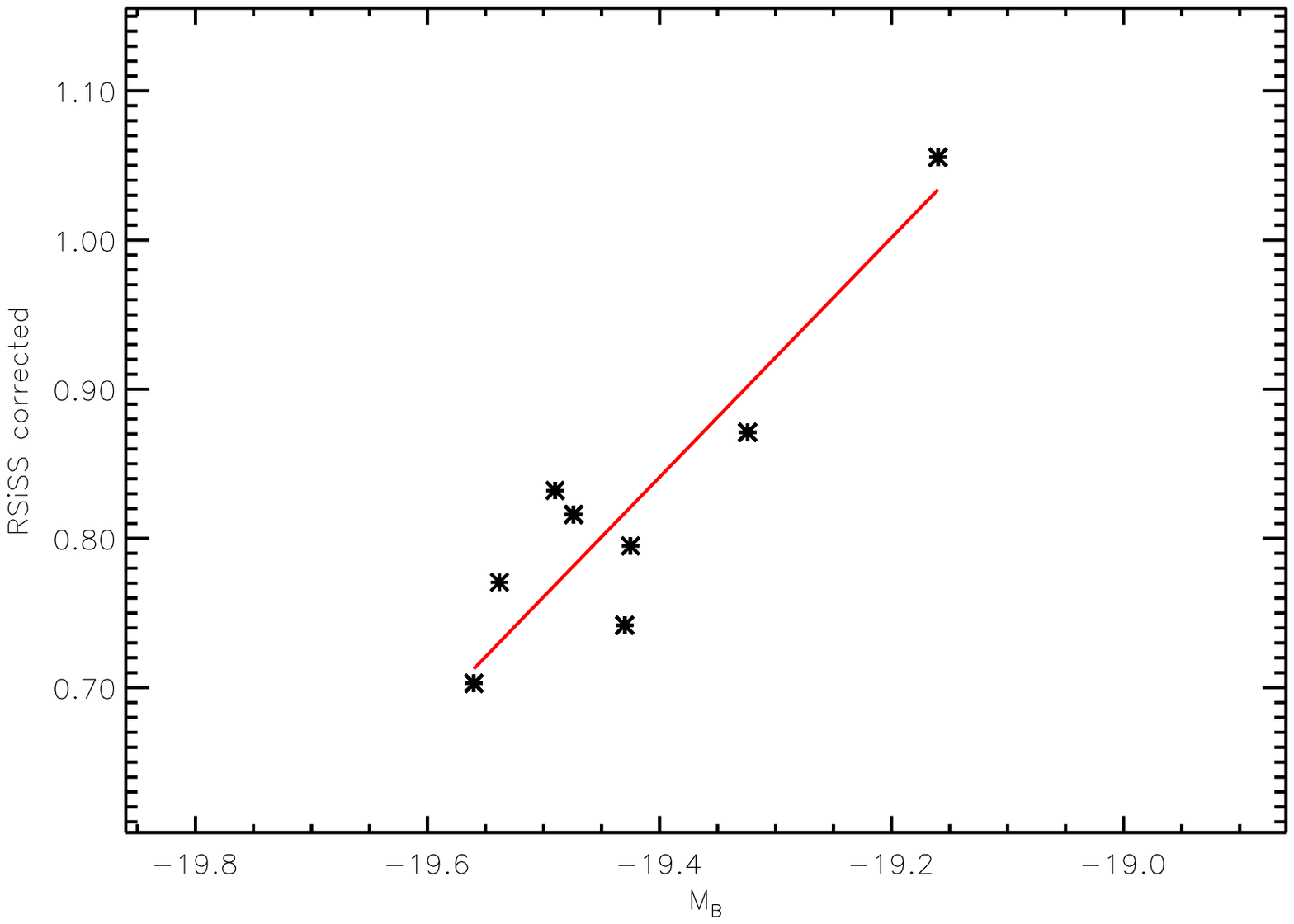}
  \caption[\RSiSu\ time evolution]{
    \RSiSuS\ time evolution (top), \RSiSuS\  corrected for the
    time dependence (bottom).} 
  \label{fig:RSiSuSrealTime}
\end{figure}

In Fig.~\ref{fig:RSiSurealTime} and Fig.~\ref{fig:RSiSuSrealTime}
we plot the time dependence of \RSiSu\ and \RSiSuS\ as well as the  
linear regression calculated on the values at $t=0$ interpolated or
extrapolated with quadratic fits. The three \supernovae with sufficient
data do not display a common time evolution, making doubtful the
existence of an universal \RSiSu\ or \RSiSuS\ time correction.
 
We again calculated the standard deviation,
the results of the time correction are summarized
in Tab.~\ref{tab:RSiSutime}.

\begin{deluxetable}{llcc}
\tablecolumns{4}
\tablewidth{0pc}
\tablecaption{\RSiSu\ and \RSiSuS\ linear regression with $t=0$
  extrapolated values\label{tab:RSiSutime}}
\tablehead{
& \colhead{$\sigma$}   &
\colhead{$a$}&
\colhead{$\sigma_{M_B}$}}
\startdata
\RSiSu\ (Case B) &         0.03 &  0.67 &    0.04 \\
\RSiSuS\ (Case B) &        0.04 &  0.80 &    0.05 \\
\RSiSu\ (Case A)  &        0.02 &  0.49 &    0.05 \\
\RSiSuS\ (Case A) &        0.04 &  0.59 &    0.06 \\
\enddata
\tablenotetext{a}{\RSiSu\ and 
\RSiSuS\ linear regression on observed \supernovae with sufficient
wavelength coverage using the $t=0$ interpolated values}
\end{deluxetable}

Correcting for the time evolution increases the \RSiSu\ and
\RSiSuS\ efficiency both by increasing the slope and decreasing the
dispersion, the only drawback is the need of at least three 
points to interpolate or extrapolate to $t=0$ with a quadratic
fit. Since there is no common trend for the three \supernovae that
were fit,
this correction remains purely empirical.

\section{Comparison with synthetic spectra}

\begin{figure}
\centering
  \includegraphics[width = 0.8\textwidth]{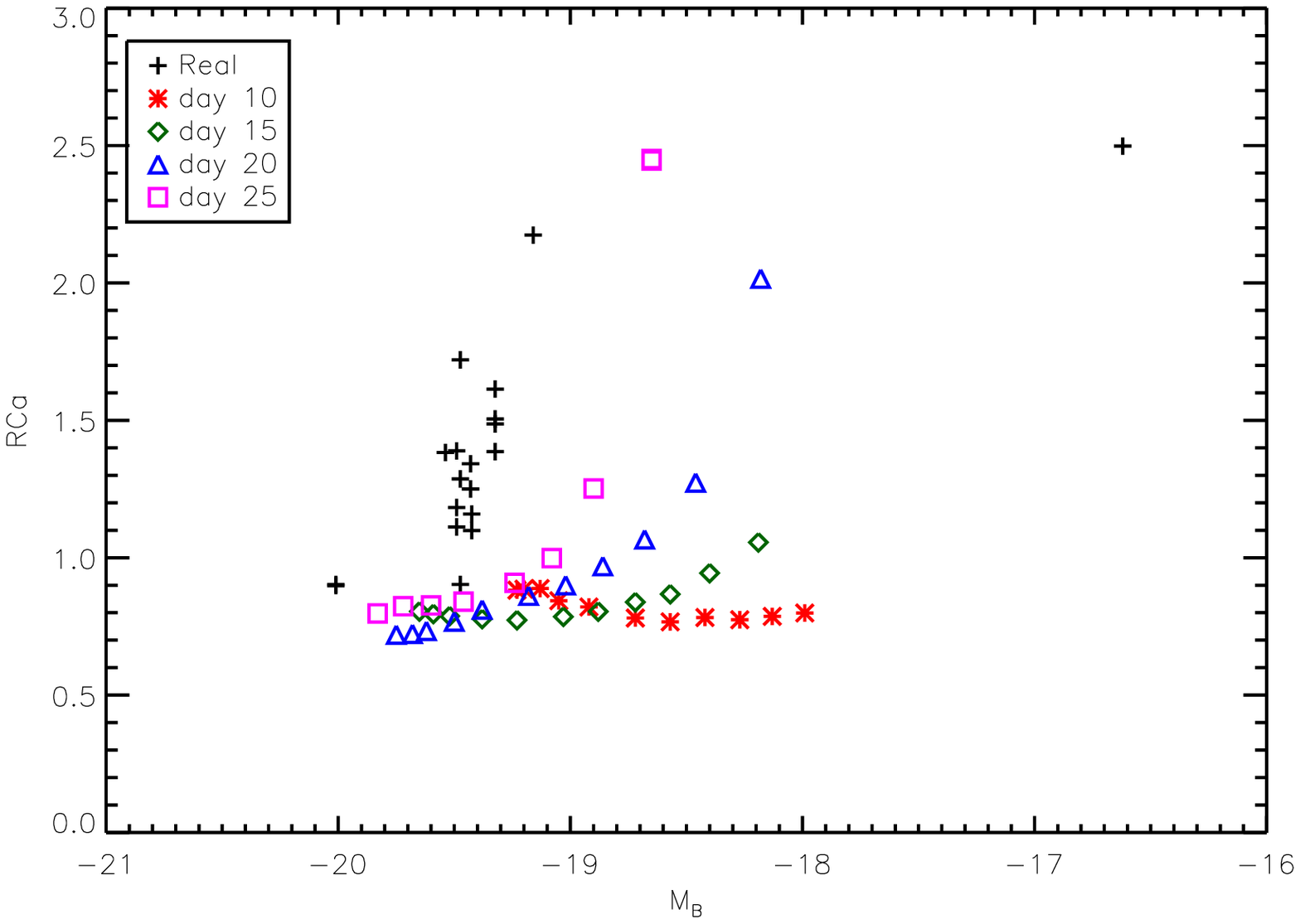}
  \caption{
    \RCa\ for Case B.
    Black: Observed \supernovae, Blue: day 20 \nomw\ \phoenix\ 
    synthetic spectra, Pink: day 25 \nomw\ \phoenix\ 
    synthetic spectra.} 
  \label{fig:RCaphx}
\end{figure}

\begin{figure}
\centering
  \includegraphics[width = 0.8\textwidth]{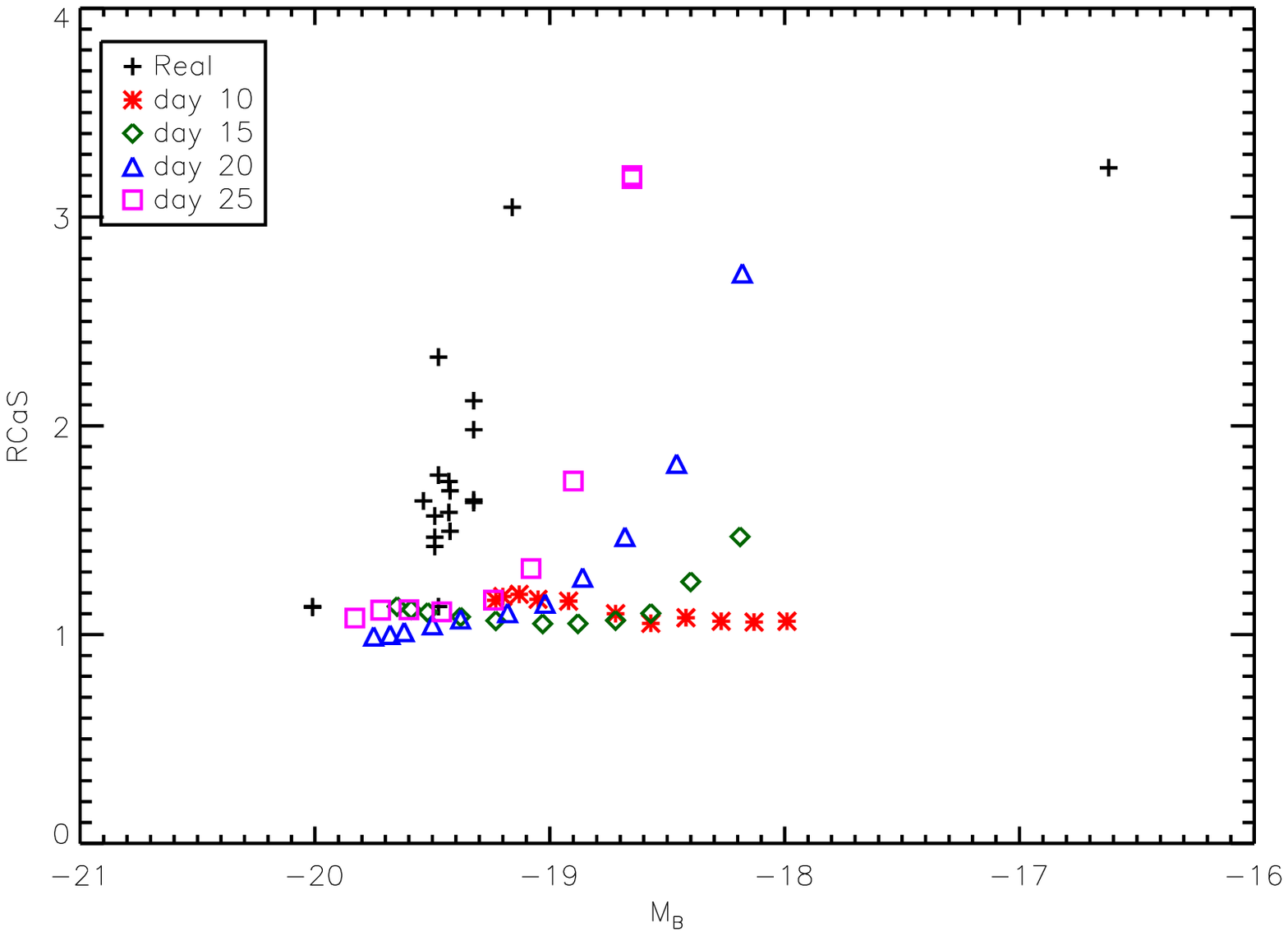}
  \caption{
    \RCaS\ for Case B.
    Black: Observed \supernovae, Blue: day 20 \nomw\ \phoenix\ 
    synthetic spectra, Pink: day 25 \nomw\ \phoenix\ 
    synthetic spectra.} 
  \label{fig:RCaSphx}
\end{figure}

In order to ascertain whether the trends that we have found for the
spectral indices can be reproduced by models, we have calculated a set
of local thermodynamic equilibrium (LTE) models  using the
generalized stellar atmosphere code \phoenix\ 
\citep{hbjcam99,hbmathgesel04} of the parameterized SN~Ia deflagration
model W7 where we varied
the total luminosity bolometric luminosity in the observer's frame. We
calculated one  
set at 20 days after explosion (the fiducial time of maximum in $V$)
and another at 25 days after explosion. We have assumed LTE for
computational expedience.  Studies by \citet{snefe296} and
\citet{whhs98} have shown that LTE is a reasonable approximation in
SNe~Ia.  
The \SiIIred\ feature is not reproduced by W7, as shown by
\citet{bbbh05}, who used extremely detailed NLTE. 
Only the general trends of the synthetic spectra can then be
expected to be matched by the real data.
The use of \phoenix\ with W7 models is described
in detail in \citet{l94d01}.

We display in Figs.~\ref{fig:RCaphx}--\ref{fig:RCaSphx} the
comparison between \RCa\ and \RCaS\ calculated on observed \supernovae
of Table~\ref{tab:snused} (all dates) and on \phoenix\ spectra. These synthetic
spectra cover a large range in $M_B$.
Since the blue magnitude increases monotonically in the \phoenix\ 
models with the bolometric magnitude, these figures indicate the
existence of a spectroscopic sequence with the bolometric
luminosity. The slope of the observed spectra is closer to that
obtained from the models $25$~d after explosion. The increase in \RCa
and \RCaS\ begins for lower blue magnitudes in the \phoenix\ synthetic
spectra than in the observed spectra.

\begin{figure}
\centering
  \includegraphics[width = 0.65\textwidth]{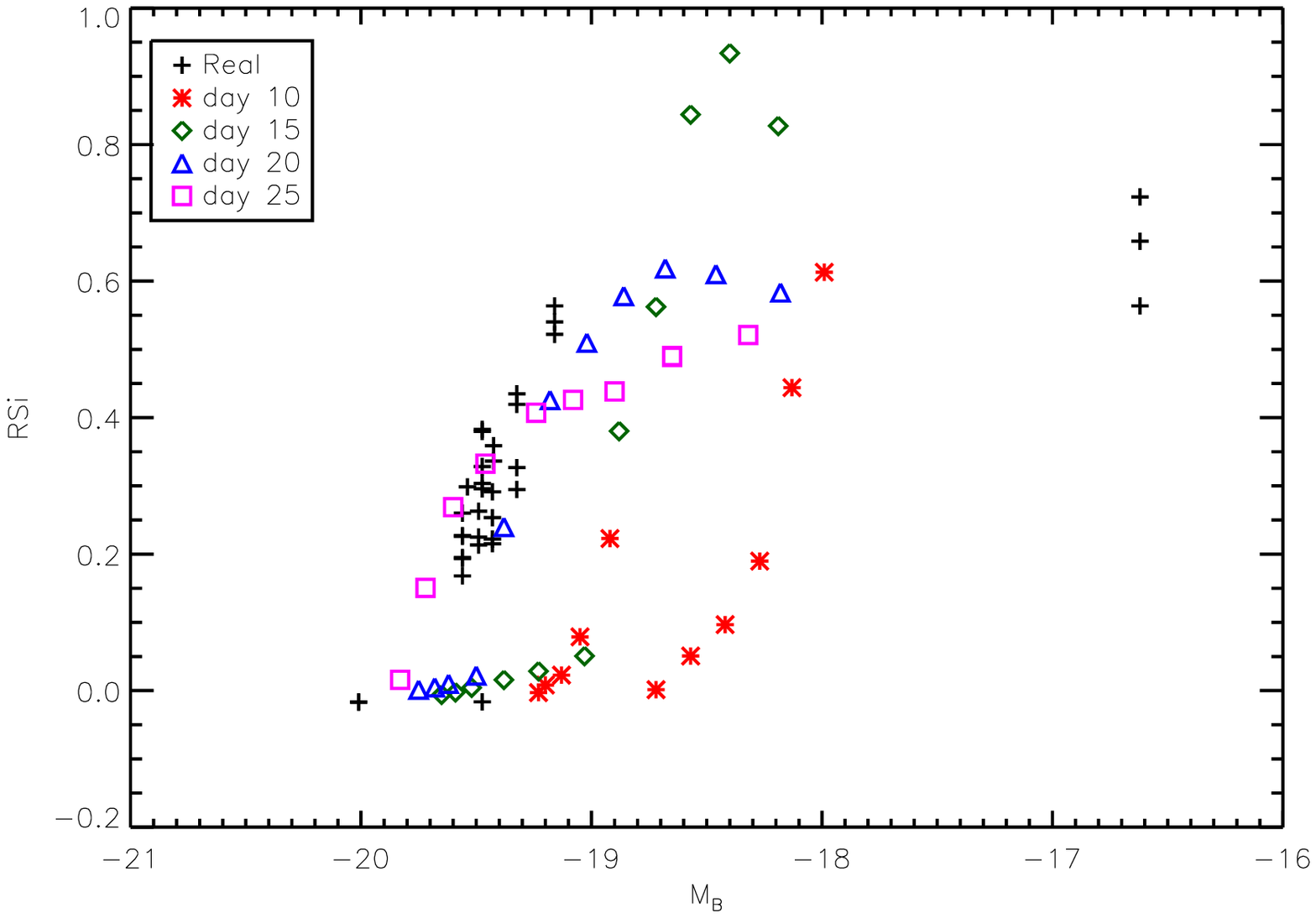}
  \includegraphics[width = 0.65\textwidth]{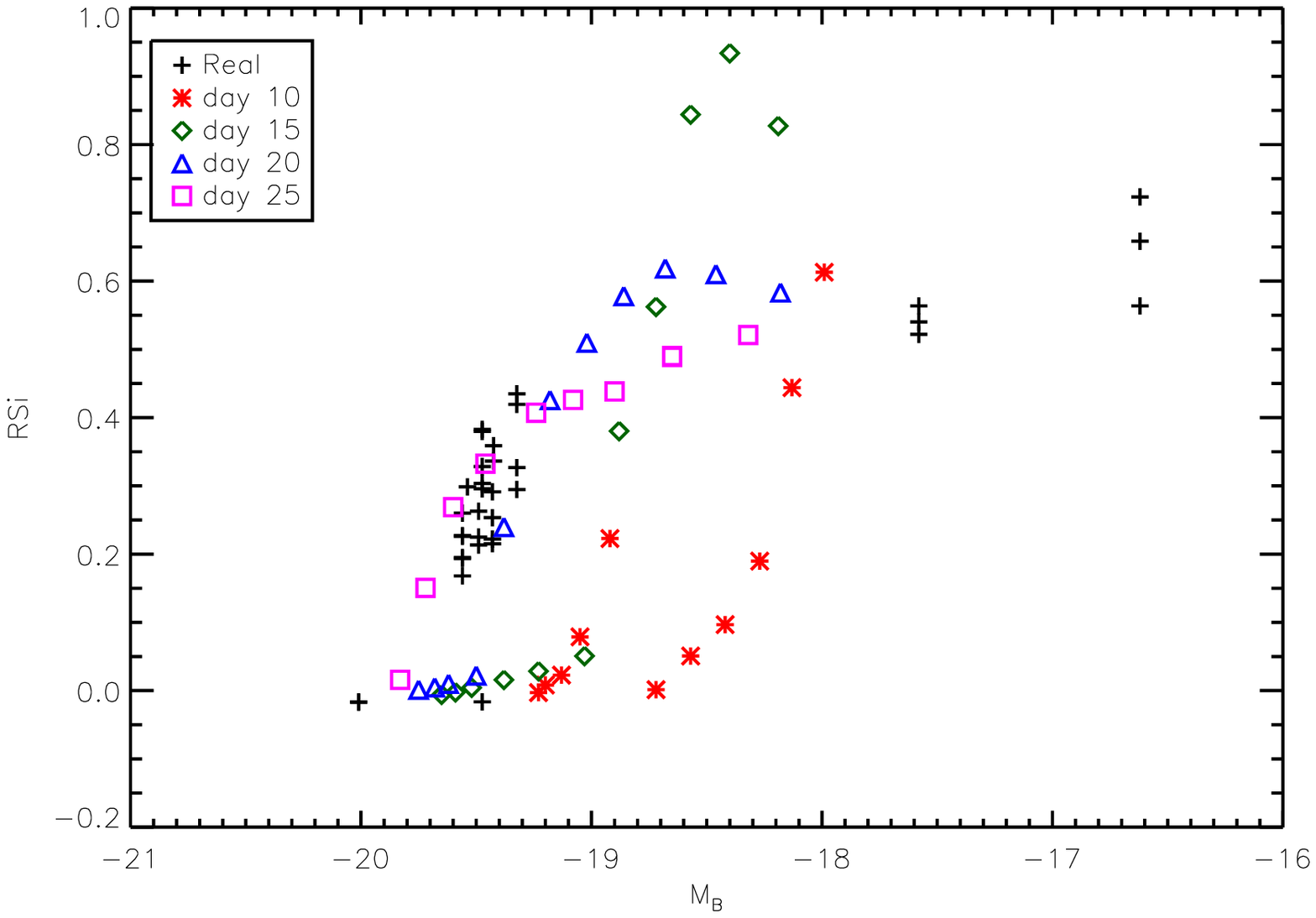}
  \caption{
    \RSi\ for $-19.16$ (top) and  $-17.58$ (bottom) SN 1986G blue
    magnitude. Black: real \supernovae. Blue: day 20 \nomw\ \phoenix\ 
    synthetic spectra. Pink: day 20 \nomw\ \phoenix\ 
    synthetic spectra. For Day 10 there is a discontinuity in \RSi
    around $M_B=-18.9$, this is real and is caused by the change in
    the formation of the \SiIIblue\ feature as discussed in Bongard
    \etal (in preparation). The very steep dependence of the various
    \RSi\ ratios as of function of $M_B$ in the range $-19.5 < M_B <
    -20$ is remarkably well reproduced by \phoenix\ at day 25. The data
    points at low luminosity suggest a smaller slope. This justifies
    the limited luminosity range for the evaluation of the slopes in \S\S~\ref{sec:rsi--rca-1}--\ref{sec:time-evolution}} 
  \label{fig:RSiphx}
\end{figure}

\begin{figure}
\centering
  \includegraphics[width = 0.8\textwidth]{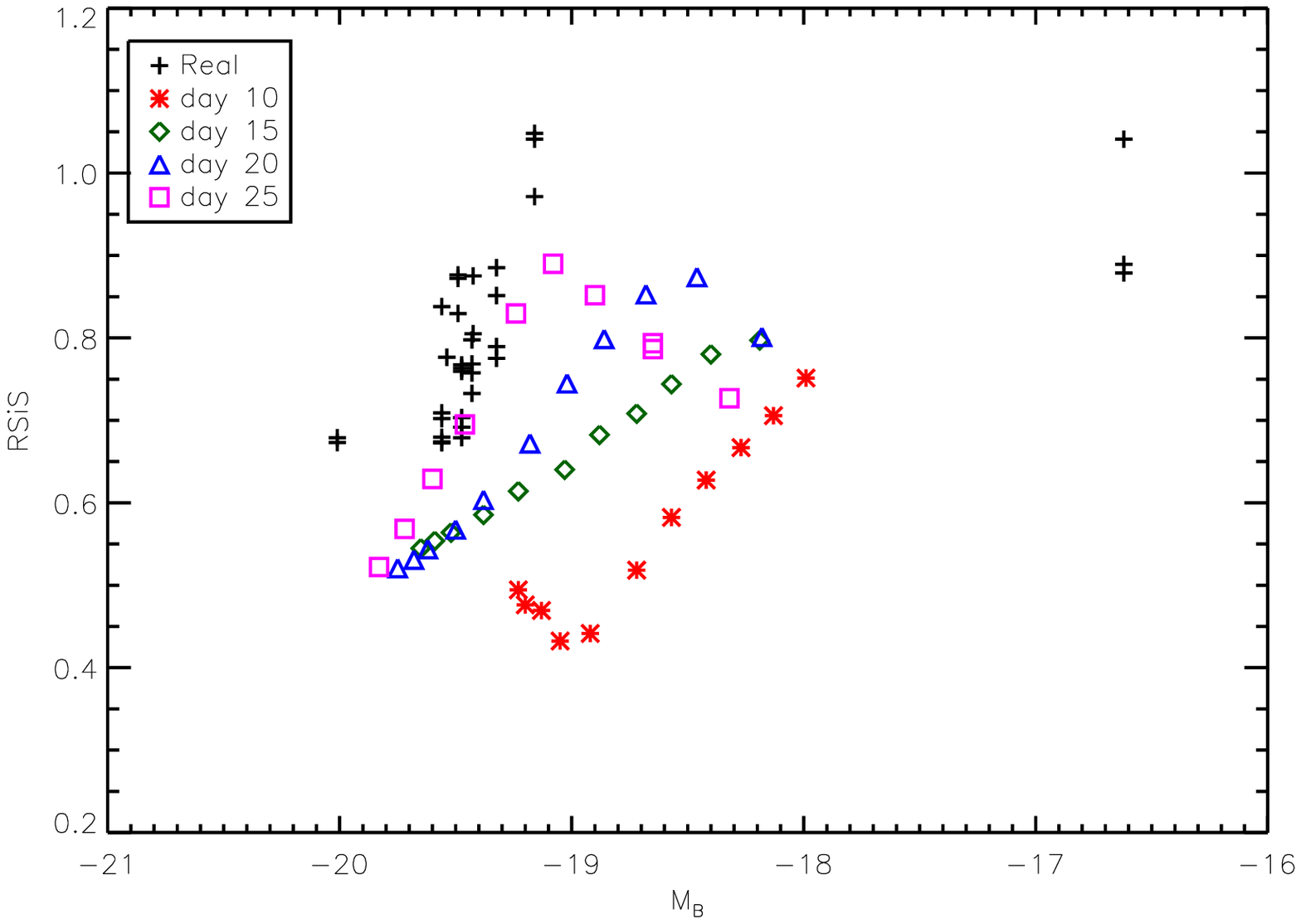}
  \includegraphics[width = 0.8\textwidth]{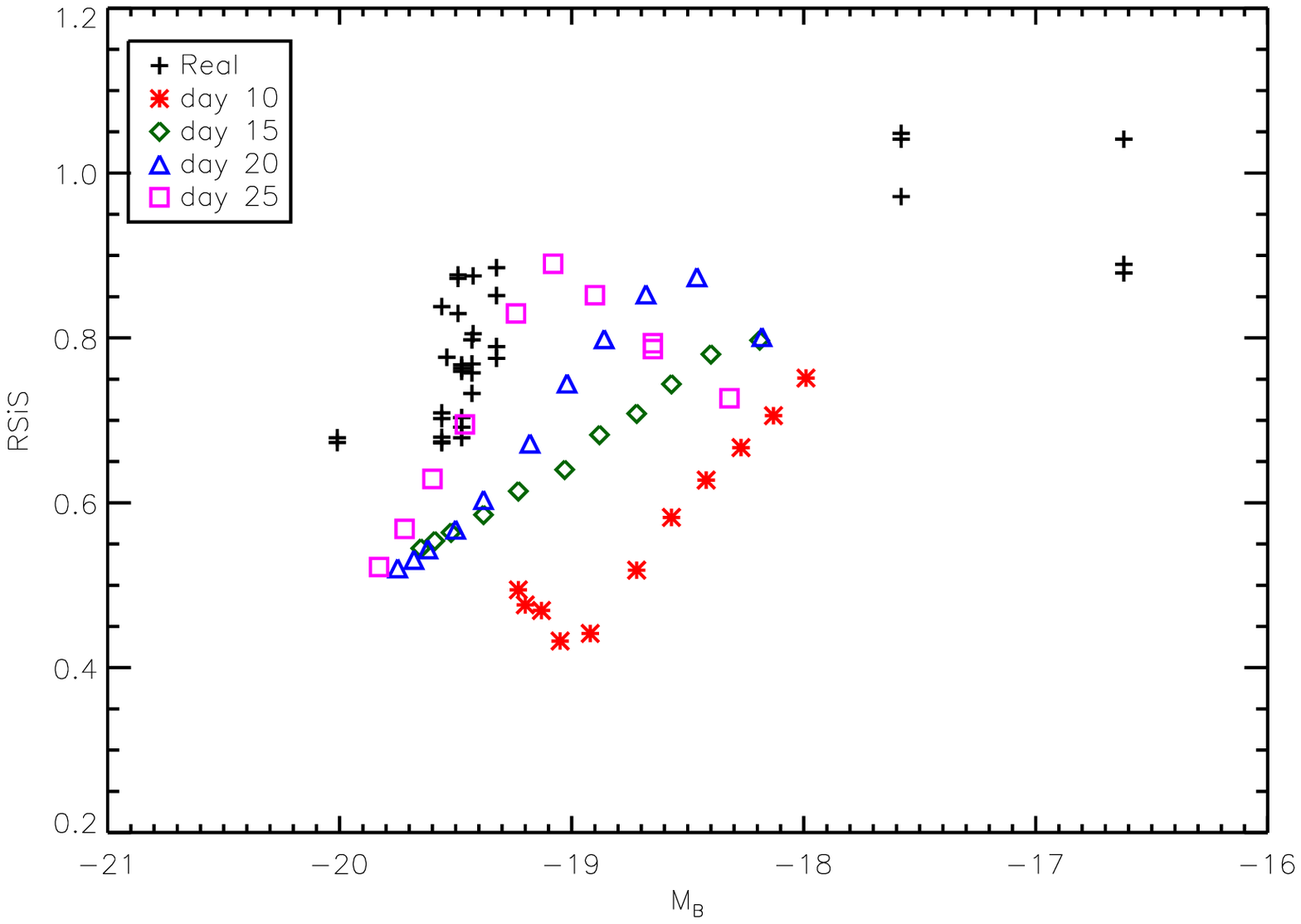}
  \caption{
    \RSiSu\ for $-19.16$ (top) and  $-17.58$ (bottom) SN 1986G blue
    magnitude. Black: real \supernovae. Blue: day 20 \nomw\ \phoenix\ 
    synthetic spectra. Pink: day 20 \nomw\ \phoenix\ 
    synthetic spectra.} 
  \label{fig:RSiSuphx}
\end{figure}

\begin{figure}
\centering
  \includegraphics[width = 0.8\textwidth]{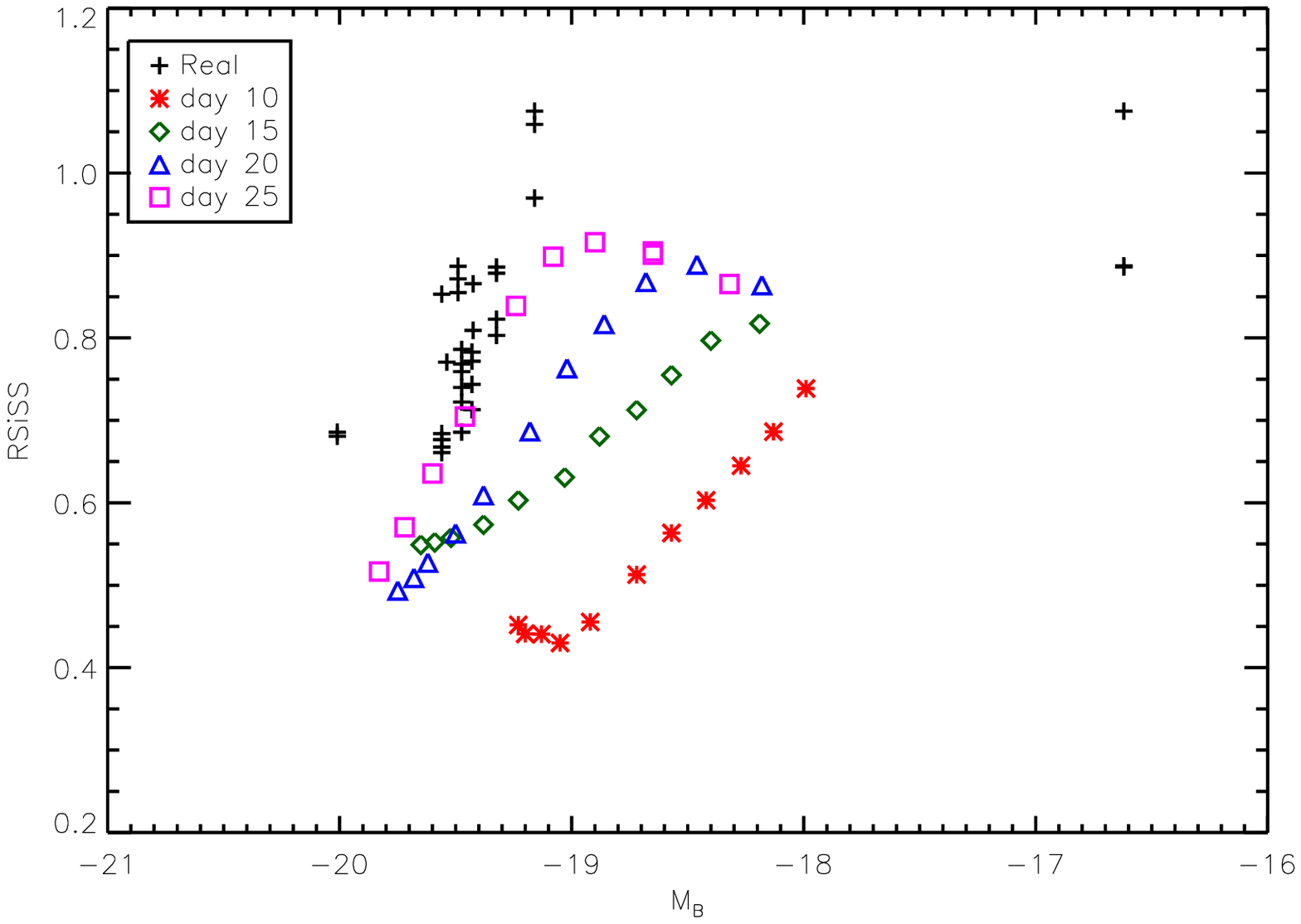}
  \includegraphics[width = 0.8\textwidth]{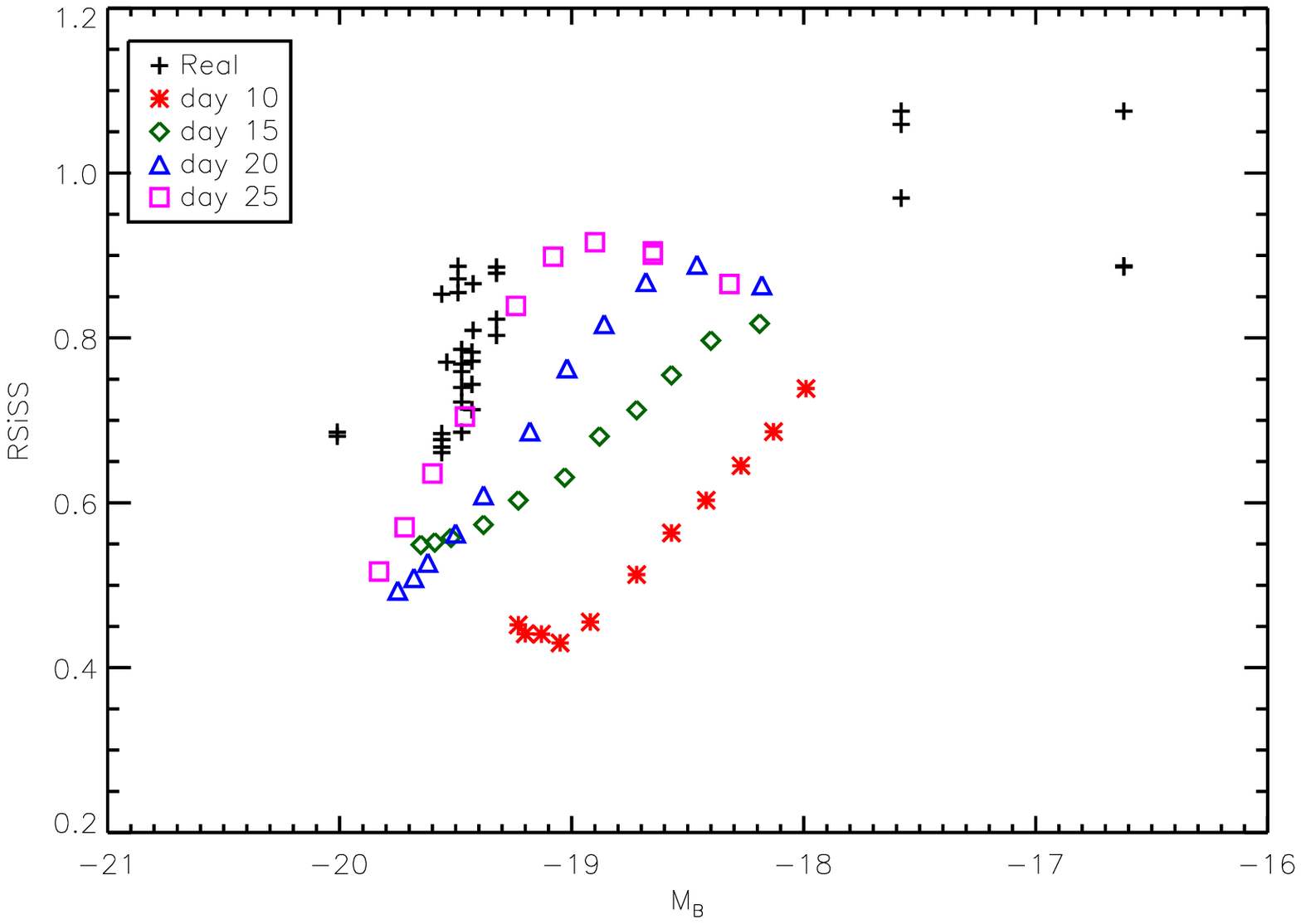}
  \caption{
    \RSiSuS\ for $-19.16$ (top) and  $-17.58$ (bottom) SN 1986G blue
    magnitude. Black: real \supernovae. Blue: day 20 \nomw\ \phoenix\ 
    synthetic spectra. Pink: day 20 \nomw\ \phoenix\ 
    synthetic spectra.} 
  \label{fig:RSiSuSphx}
\end{figure}

Figs.~\ref{fig:RSiphx}--\ref{fig:RSiSuSphx} display the same synthetic
and observed spectra for 
\RSi, \RSiSu, and \RSiSuS. Again, the existence of the spectroscopic
sequence is clear in both the observed and the synthetic
spectra. Moreover, while the general trend is only qualitatively similar
for \RCa\ and \RCaS, \RSi\ and to a lesser extent \RSiSu\ and \RSiSuS
show that these ratio behave in the same way with respect to the blue
magnitude evolution for synthetic as well as for real spectra.

\section{Use of line ratios in the \snap\ context}
\label{sec:use-these-line}

In order to estimate the luminosity measure accuracy of these spectral 
indicators in a cosmological context, we  simulated
\snap\ exposures \citep{snap_detf_wp05} at different redshifts. We used
them to estimate the 
accuracy on the determination of the mean luminosity of ``Branch
normal'' \supernovae, which is the value needed to devise the Hubble 
diagram and the determination of cosmological parameters.

\subsection{\snap\ simulator}
\label{sec:snap-resol-progr}

\subsubsection{The \snap\ spectrometer}
\label{sec:resolution-noise}

To simulate the \snap\ spectrometer, we assumed  $50\%$ optical
and  $70\%$ CCD efficiencies \citep{kimetal04}. We considered the
mirror surface to be 
$ \pi~ m^{2}$ (i.e. a 2$m$ mirror diameter)
and implemented the re-binning of the  
\supernovae spectra to \snap\ binning in the observer rest frame.
The red and blue channel resolution of \snap\ we used are 
summarized in Table~\ref{tab:snapB} (A.~Ealet, private communication;
\citeauthor{ealetetal03} [\citeyear{ealetetal03}]).   

\begin{deluxetable}{cccccccc}
\tablecolumns{8}
\tablewidth{0pc}
\tablecaption{\snap\ blue and red channel resolution\label{tab:snapB}}
\tablehead{\colhead{BLUE} &&&&\colhead{RED}\\
\colhead{$\lambda$ (\AA)}    &   \colhead{$\frac{\lambda}{\Delta\lambda}$}   &
\colhead{$\lambda$ (\AA)} &  \colhead{$\frac{\lambda}{\Delta\lambda}$}& 
\colhead{$\lambda$ (\AA)}    &   \colhead{$\frac{\lambda}{\Delta\lambda}$}   &
\colhead{$\lambda$ (\AA)}    &
\colhead{$\frac{\lambda}{\Delta\lambda}$}
}
\startdata
  4000 & 295. & 7500 & 95.  \\ 
  4500 & 235. & 8000 & 90.  \\   
  5000 & 185. & 8500 & 85.  \\   
  5500 & 155. & 9000 & 80.  \\   
  6000 & 130. & 9500 & 78.  \\   
  6500 & 115. & 10000& 75.  \\   
  7000 & 105. &      &      \\

&&&&    10000.  & 78.  & 14000.  & 86.  \\   
&&&&    10500.  & 77.  & 14500.  & 90.  \\   
&&&&    11000.  & 77.  & 15000.  & 93.  \\   
&&&&    11500.  & 78.  & 15500.  & 97.  \\   
&&&&    12000.  & 79.  & 16000.  & 100. \\   
&&&&    12500.  & 80.  & 16500.  & 105. \\   
&&&&    13000.  & 82.  & 17000.  & 110. \\   
&&&&    13500.  & 84.  &         &      \\
\enddata
\tablenotetext{a}{The resolution is taken from A. Ealet (private
  communication).}
\end{deluxetable}

The separation between the two spectrograph channels  is not yet
precisely defined, therefore we treated the spectra as if the red and
the blue detector 
properties were the same. Since the edges of the spectral range
are  harder to calibrate,  their actual location can
impact the final results.

\subsubsection{The simulated \supernovae}
\label{sec:simulated-supernovae}

In order to simulate \snap\ exposures, we calibrate each supernova  to
its absolute blue magnitude and then redshift it to the desired
$z$.  

\begin{equation}
  \label{eq:17}
  N_{\gamma}(z)=N_{\gamma}(0) (1+z)
  \left(\frac{d_{0}}{d_{\mathrm{L}}}\right)^{2} 
\end{equation}

Eq.~(\ref{eq:17}) shows how the number of photons per unit time and
per unit area in the redshifted $B$ band in the supernova rest frame
at redshift $z$, 
$N_{\gamma}(z)=\int_{B(z)} \frac{dN_{\gamma}(z)}{d\lambda}d\lambda$,
is related to $N_{\gamma}(0)=\int_{B(0)}
\frac{dN_{\gamma}(0)}{d\lambda}d\lambda$,  
the number of photons per unit of time and of surface in the B band at
rest in the supernova rest frame for a supernova at 10~pc. 

The flux dilution factor for a supernova at $z$ is
$\frac{d_{0}^{2}}{d_{\mathrm{L}}^{2}}$ where $d_{0}=10$~pc, but the
reddening factor $1+z$ included in $d_{\mathrm{L}}$ gets cancelled when
photons are counted in redshifted spectral bands. 
We also introduced electronic and statistical noise using the values
given in
Table~\ref{tab:noise}. 

\begin{deluxetable}{ll}
\tablecolumns{2}
\tablewidth{0pc}
\tablecaption{Noise per pixel per 2000
    s.\label{tab:noise}}
\tablehead{}
\startdata
    readout noise/pixel   & $\sqrt{2} \cdot 5 e$\\
    slow drift contribution     &  $7 \cdot 1.2 e$ \\
    Leakage current fluct. for 2000s   &  $6e$ \\
    Total electronic noise per pixel:            & $10e$ \\
    Presence of a cosmic ray ($20\%$ of cases) & $12e$ \\
    Total Electronic noise per $\Delta\lambda$ (3 spatial pixels) & $17e$ \\
\enddata
\end{deluxetable}

The electronic noise for each $2000 s$ exposure was calculated using
\begin{equation}
  \label{eq:noise}
  n_{electronic}^{2}=17^{2} \cdot \frac{t}{2\times 10^{3}}
\end{equation}
and assumed to be a Gaussian  with zero mean. The Poisson noise of
the signal was also included for each of the $2000$~s exposures needed
to complete the total exposure time.

\begin{deluxetable}{ccc}
\tablecolumns{3}
\tablewidth{0pc}
\tablecaption{\snap\ simulator $N_{\gamma}~s^{-1}$ in Bessel B
\label{tab:nphot}}
\tablehead{
\colhead{$z$}    &   \colhead{Blue filter ($N_{\gamma}~s^{-1}~cm^{2}$)}   &
\colhead{Exposure  Time}}
\startdata
   $1.0$ & $4.32$ & $5980 s$ \\
   $1.5$ & $1.75$ & $22700 s$ \\
   $1.7$ & $1.3$ & $36000 s$ \\
\enddata
\tablenotetext{a}{The 
    number of photons does not take into account the factor $0.5*0.7$
    from the CCD efficiency. It is calculated here for a $M_{B}=
    -19.42$ \SNIa.}
\end{deluxetable}

In Table~\ref{tab:nphot} we display the number of photons per second
expected in the B Bessel filter for a \SNIa of absolute blue magnitude
$M_{B}= -19.42$, not accounting for the CCD and optical
transmission. The exposure time we use have been assumed to follow a
$(1+z)^{6}$ power law for $1 < z < 1.7$ and an exposure time of $10$ hours for
$z=1.7$, chosen in order to maintain a constant signal to noise as $z$
increases (A.~Ealet, private communication).  
   
\subsection{Results  at  $z=1.5$}
\label{sec:results}

\begin{deluxetable}{ccc}
\tablecolumns{3}
\tablewidth{0pc}
\tablecaption{\label{tab:snSnap}\Supernovae 
    simulated in the \snap\ context.}
\tablehead{
\colhead{Supernova}    &   \colhead{Date (wrt maximum)}   &
\colhead{$M_\textrm{B}$}}
\startdata
SN 1991T & $0$ & $-20.01$ \\
  SN 1998aq & $0, 1, 2$ & $-19.56$ \\
SN 1981B & $0$ & $-19.54$ \\
  SN 1998bu & $ -2, -1$ & $-19.49$ \\
SN 1994D & $-3$ & $-19.47$ \\
SN 1996X  & $-2, 0, 1$ & $-19.43$ \\            
SN 1989B & $-1$ & $-19.42$ \\
SN 1992A & $-1$ &  $ -19.32$ \\
SN 1986G & $-1$ & $-17.58$ or $-19.16$ \\
SN 1991bg & $0$ & $-16.62$ \\
\enddata
\end{deluxetable}

\begin{figure}
\centering
  \includegraphics[width = 0.8\textwidth]{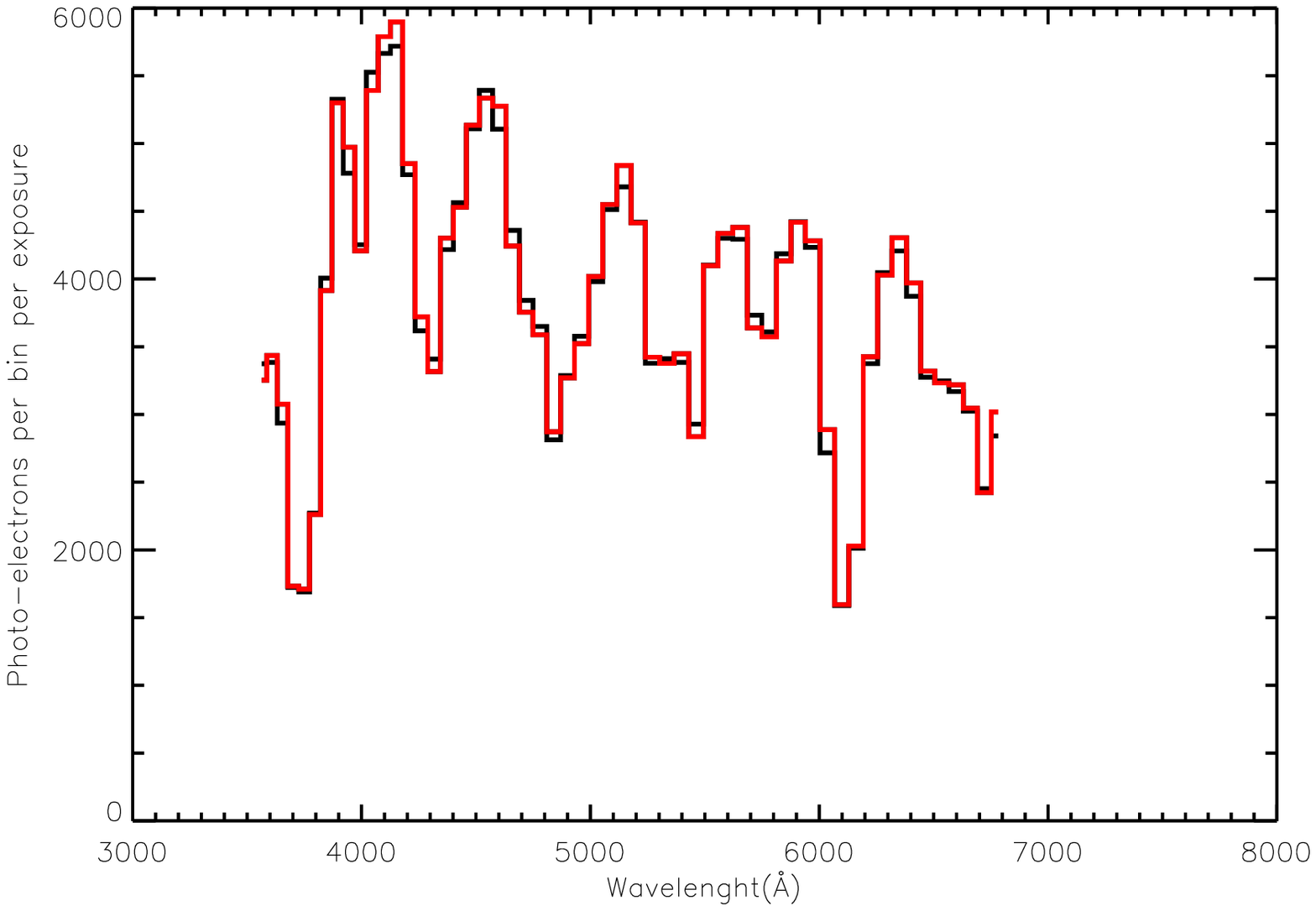}
  \includegraphics[width = 0.8\textwidth]{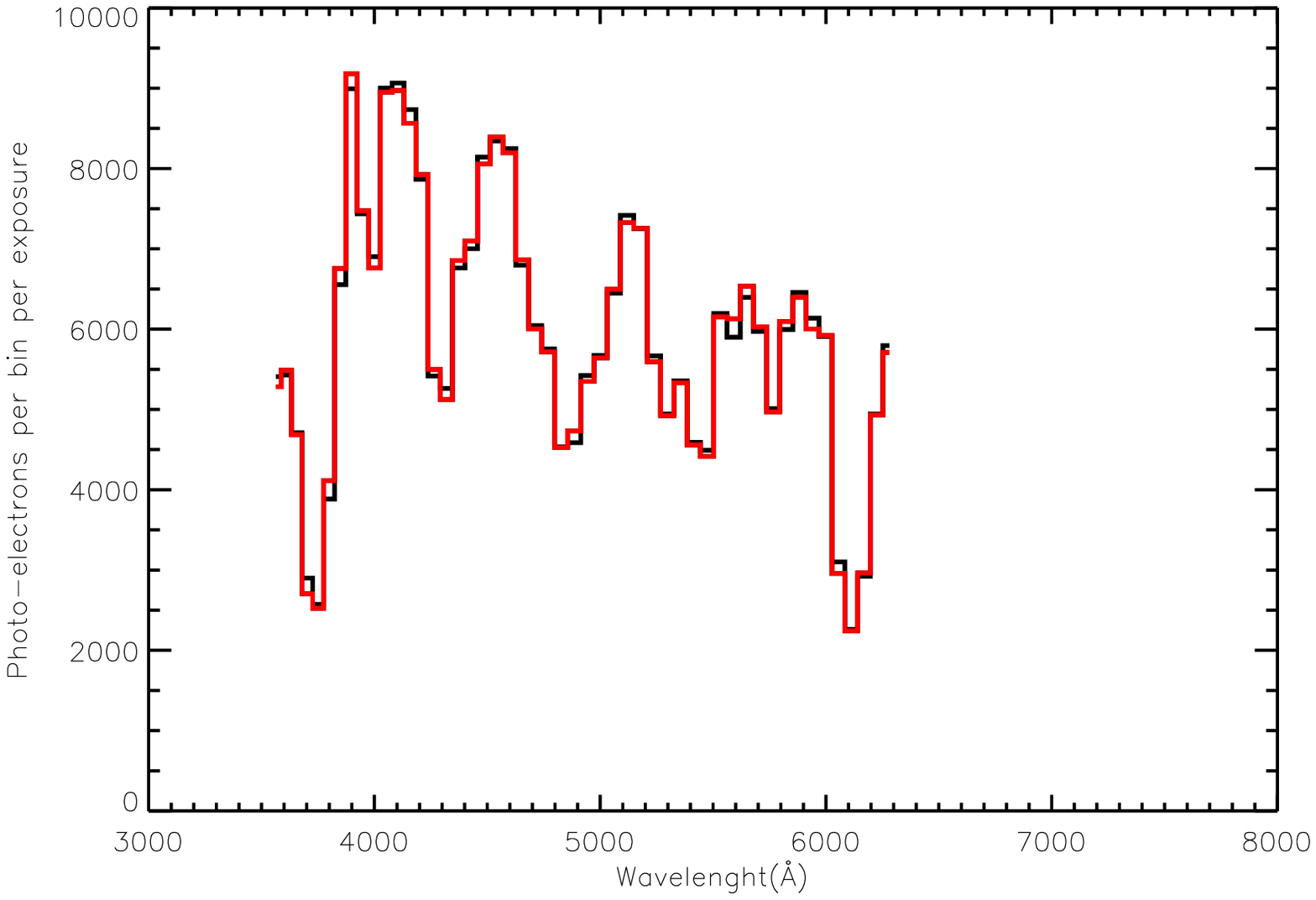}
  \caption[\snap\ simulations]{\snap\ SN 1992A
      $z=1.5$ simulated exposure (top), and $z=1.7$ (bottom). Black
      curves are only the supernova, red curves include  
    Poisson and electronic noise.}  
  \label{fig:snapSimPic}
\end{figure}

We display in Fig.~\ref{fig:snapSimPic} the
number of photons per bin simulated obtained for the standard
supernova SN~1992A at $z=1.5$ and $z=1.7$, respectively. This  
quantity differs from the usual flux per \AA, which explains the  
flatter shape of the spectrum as there is a 
difference of $1/\lambda$  between them. For $z=1.7$, the wavelength
coverage will be too small to allow the calculation of \RSi\ and
\RSiSu, although it would be accessible to JEDI (the proposed
spectroscopic wavelength coverage for JEDI is 0.8--2~$\mu$m). 

We simulated 100 exposures for every supernova listed in
Table~\ref{tab:snSnap} and calculated the spectral indicators
previously defined for each of them and found the mean and standard
deviation (see Tables~\ref{tab:RSi1}, \ref{tab:RSi2}, \ref{tab:RSiSu1},
and~\ref{tab:RSiSu2}). From this we calculate $\sigma_{M_B}$ assuming
that the slope of the linear regression is known perfectly.
In order to concentrate on the accuracy of the spectroscopic
indicators in \snap, and since at redshifts  of $z \approx 1$ time 
dilatation doubles the evolution time of the light curve, we assumed
that the spectra will be taken almost at  maximum light. We
thus included in our sample only the supernova spectra that were the
close to the light curve maximum.

\subsubsection{\RSi}
\label{sec:rsi}

We list in Table~\ref{tab:RSi1} the luminosities measured for the
simulated \supernovae without any selection cut on the simulated \RSi\ 
values for Case A. The RMS dispersion due to the noise is displayed as well as
the mean \RSi\ used to calculate the blue magnitude. In this first
table we used the linear regression calculated 
with \RSi\ values extrapolated to $t=0$ when possible. 
 Table~\ref{tab:RSi2} lists the same
values but for the linear regressions corresponding to Case B.
The large discrepancies observed for SN~1991T and SN~1991bg between
the inferred (using \RSi) luminosity and the actual value suggests
that the linear regression is invalid at low and high magnitudes, we
discuss this further below; however, no firm conclusion can be drawn
from the single cases in our current sample. 


In order to
estimate the mean accuracy for JDEM we add the values of $\delta
M_{B}$ listed in the last column of the table in quadrature and divide
by $N-2$. We lose two degrees of freedom by finding the mean and the
slope with the same sample of \supernovae in
Tables~\ref{tab:RSi1}--\ref{tab:RSi2} (and then we assume that the
slope used is known perfectly); these values are 
listed in Table~\ref{tab:snapressum}.
Red,
dim SN~1991bg-like SNe~Ia will be strongly selected against due to
Malmquist bias in deep cosmological surveys, therefore
we shall never consider them in our analyses (in Case A we include
SN~1986G in the 91bg-like class). Similarly SN~1991T-likes would also
not be used for precision cosmology, but we quote results both
including and excluding 91T-likes.
In Case B where blue magnitude of SN 1986G  is $M_B=-19.16$ as is favored by
\citet{reindl05}.  the mean accuracy on
the blue magnitude becomes $\approx 0.17 (0.66)$  blue magnitude excluding 
(including) 91T-likes. 


\begin{deluxetable}{lcccccc}
\tablecolumns{7}
\tablewidth{0pc}
\tablecaption{\RSi\ (Case A)\label{tab:RSi1}}
\tablehead{
 \colhead{SN name}  &   \colhead{mean}   & \colhead{
   $\sigma_{\textrm{\RSi}}$} & \colhead{Inferred $M_{B}$} &
 \colhead{$\sigma_{M_{B}}$} & \colhead{$\delta M_{B}$}}
\startdata
91T &       1.16 &      0.81 & -16.84 &        2.34 &       -3.17 \\
98aq &      0.18 &     0.08 &  -19.66 &       0.23 &  0.10 \\
81B &      0.33 &     0.09 &   -19.24 &       0.26 &      -0.3 \\
98bu &      0.17 &     0.05 &   -19.70 &       0.16 &       0.21 \\
94D &      0.29 &     0.09 &   -19.35 &       0.27 &      -0.12 \\
96X &      0.27 &     0.09 &   -19.42 &       0.27 &    -0.01 \\
89B &      0.36 &     0.09 &   -19.14 &       0.33 &      -0.27 \\
92A &      0.26 &     0.11 &   -19.44 &       0.36 &       0.12 \\
86G &      0.32 &      0.16 &   -19.26 &       0.47 &        1.68 \\
91bg &      0.56 &      0.37 & -18.58&       1.09 &        1.96 \\
\enddata
\tablenotetext{a}{$\delta M_{B}$ is the difference between the true
  value and that inferred using our correlation with the spectral
  index \RSi}
\end{deluxetable}

\begin{deluxetable}{lcccccc}
\tablecolumns{7}
\tablewidth{0pc}
\tablecaption{\RSi\ (Case B)\label{tab:RSi2}}
\tablehead{
 \colhead{SN name}  &   \colhead{mean}   & \colhead{
   $\sigma_{\textrm{\RSi}}$} & \colhead{Inferred $M_{B}$} &
 \colhead{$\sigma_{M_{B}}$} & \colhead{$\delta M_{B}$}}
\startdata
91T &       1.16 &      0.81 &  -18.33 &     1.02 &     -1.68 \\
98aq &      0.18 &      0.08 &  -19.56 &     0.10 &      0.00 \\
81B &       0.33 &      0.09 &  -19.37 &     0.12 &     -0.16 \\
98bu &      0.17 &      0.05 &  -19.58 &     0.07 &      0.09 \\
94D &       0.29 &      0.09 &  -19.43 &     0.12 &     -0.04 \\
96X &       0.27 &      0.09 &  -19.46 &     0.12 &      0.02 \\
89B &       0.36 &      0.09 &  -19.34 &     0.15 &     -0.08 \\
92A &       0.26 &      0.11 &  -19.47 &     0.16 &      0.14 \\
86G &       0.32 &      0.16 &  -19.39 &     0.21 &      0.22 \\
91bg &      0.56 &      0.37 &  -19.08 &     0.48 &      2.47 \\
\enddata
\tablenotetext{a}{$\delta M_{B}$ is the difference between the true
  value and that inferred using our correlation with the spectral
  index \RSi}
\end{deluxetable}


Table~\ref{tab:snapressum} summarizes the total error expected in
\snap. We have
calculated the total error by adding the ``intrinsic $\sigma_{M_B}$''
(obtained from the linear regression fit) in quadrature with the
$\sigma_{M_B}$ estimated above due to noise from our simulations. We
note that both error estimates include errors due to the same
systematics and thus our estimate of the total error using Method 1 is
quite conservative

We also note that our estimate of ``the intrinsic $\sigma_{M_B}$'' is
strongly dependent on the slope of the linear regression and thus our
restriction to Branch normals in obtaining the slope is
optimistic. Nevertheless, the results of the synthetic spectral calculations
bolster our view that the linearity is more pronounced in the Branch
normal region.
For \RSi,  using the $t=0$ extrapolated linear
regression instead of the one without correction increases the slope
enough to give a significant effect in the less favored Case A, by 
$\approx 0.1$ blue magnitude. Thus, to take full
advantage of this luminosity measure, we need spectra as close to
to maximum light as possible, which will be feasible at
$z=1.5$ since the elapsed time between $-1$ and $+1$ days around
maximum is $2.5 \times 2 = 5$ days and since \snap\ will try to obtain
spectra near peak to maximize signal to noise,  the proposed JEDI
mission will easily fulfill this goal since they
will obtain multiple spectra every $\approx 7$~days.

\begin{deluxetable}{ccccccccccccc}
\tablecolumns{13}
\tablewidth{0pc}
\tablecaption{Number of
    \supernovae in redshift bins for \snap. \label{tab:snz}
}
\tablehead{}
\startdata
 z  & 0.6 & 0.7 & 0.8 & 0.9 & 1.0 & 1.1 & 1.2 & 
1.3 & 1.4 & 1.5 & 1.6 & 1.7 \\
\# of SN & 150 & 171 & 183 & 179 & 170 & 155 & 142 
& 130 & 119 &  107& 94 & 80 \\
\enddata
\tablenotetext{a}{These results were obtained from \citep{aldering_snap04}}
\end{deluxetable}

As shown in Table~\ref{tab:snz}, \snap\ is expected to find $\approx
107$ \SNeIa at $z=1.5$.  The accuracy expected on the luminosity
measure with \RSi\ would then be statistically improved by a factor of
$\sqrt{107}\approx 10$ (although as discussed above this floor will
truly be determined by the systematic error). The mean \SNeIa blue
magnitude at $z=1.5$ that is the quantity of cosmological interest,
will be determined with an accuracy of $\approx (0.2)/10 \approx
0.02$blue magnitude in Case A, excluding 91T-likes. Clearly in Case A
if 91T-likes are included the method is not useful. It seems likely
that 91T-likes 
will be identifiable by JDEM, so the smaller error is more reasonable.
In Case B where its blue magnitude is $-19.16$, the mean accuracy
would be $ \approx (0.01-0.07)$, depending on whether 91T-likes are
included. Thus  \RSi\ appears to be a good candidate for a
supplementary tool to determine the equation of state of the universe
using \supernovae.

\subsubsection{\RSiSu}
\label{sec:rsisu}

In Tables~\ref{tab:RSiSu1}, and~\ref{tab:RSiSu2}
we list the results of our simulations for both
Cases A and B. Table~\ref{tab:snapressum} shows that \RSiSu\ is clearly
a superior luminosity indicator in that the total error is nearly
constant at $\la 0.2 M_B$ for all cases.
This makes this ratio appear to be an excellent secondary probe of \SNeIa
blue magnitudes and to independently test evolutionary effects.
   
In Case B, where 91T-likes are rejected, the preferred case, the mean
blue magnitude accuracy becomes $\approx 0.10$ blue magnitude,
including 91T-likes increases the dispersion to only $0.16$ which is
similar to the $\approx 0.2$ blue magnitudes dispersion of the \SNeIa
after stretch factor correction. We see that \RSiSu\ is more robust
than \RSi\ as the total error remains below $0.20$ blue magnitudes for
any of our assumptions. 


We found that \RSiSuS\ did not improve the results. This is due to the
large wavelength binning in the \snap\ design that already effectively
transforms \RSiSu\ into an ``integral'' ratio compared to the $10$\AA\
binning we previously used, as can be seen from the low dispersion due
to Poisson noise($>5\%$). Thus \RSiSuS\ results are almost identical to
those of \RSiSuS, sometimes even a bit worse because the larger \snap\
binning can include more of the \FeII $\approx 5600$\AA\ peak into the
integral, reducing its quality as a luminosity indicator.

With more \supernovae added to our analysis it may be possible to
include both 1991bg-like events into our calibration
for the spectral indices.

\begin{deluxetable}{lcccccc}
\tablecolumns{7}
\tablewidth{0pc}
\tablecaption{\RSiSu\ Case A\label{tab:RSiSu1}}
\tablehead{
 \colhead{SN name}  &   \colhead{mean}   & \colhead{
   $\sigma_{\mathrm{\RSiSu}}$} & \colhead{Inferred $M_{B}$} &
 \colhead{$\sigma_{M_{B}}$} & \colhead{$\delta M_{B}$}}
\startdata
91T &      0.66 &     0.03 &      -19.71  &      0.07 &     -0.30 \\
98aq &      0.69 &     0.03 &     -19.64  &      0.06 &      0.08 \\
81B &      0.78 &     0.04 &      -19.47  &      0.07 &     -0.07 \\
98bu &      0.86 &     0.05 &      -19.31 &      0.11 &      -0.18 \\
94D &      0.77 &     0.04 &      -19.48  &      0.07 &      0.01 \\
96X &      0.74 &     0.04&       -19.54  &      0.08 &       0.11 \\
89B &      0.83 &     0.06 &      -19.37  &      0.13 &     -0.05 \\
92A &      0.89 &     0.07 &      -19.24  &      0.15 &     -0.08 \\
86G &       1.06 &     0.07 &     -18.89  &      0.15 &        1.31 \\
91bg &      0.94 &      0.21 &    -19.15  &      0.43 &        2.52 \
\enddata
\tablenotetext{a}{$\delta M_{B}$ is the difference between the true
  value and that inferred using our correlation with the spectral
  index \RSiSu}
\end{deluxetable}

\begin{deluxetable}{lcccccc}
\tablecolumns{7}
\tablewidth{0pc}
\tablecaption{\RSiSu\ Case B\label{tab:RSiSu2}}
\tablehead{
 \colhead{SN name}  &   \colhead{mean}   & \colhead{
    $\sigma_{\textrm{\RSiSu}}$} & \colhead{Inferred $M_{B}$} &
 \colhead{$\sigma_{M_{B}}$} & \colhead{$\delta M_{B}$}}
\startdata
91T &       0.66 &     0.03 &        -19.64 &      0.05 &     -0.36 \\
98aq &       0.69 &     0.03 &       -19.60 &      0.04 &      0.04 \\
81B &       0.78 &     0.04 &        -19.48 &      0.05 &     -0.06 \\
98bu &       0.86 &     0.05 &       -19.36 &      0.08 &     -0.13 \\
94D &       0.77 &     0.04 &        -19.49 &      0.05 &      0.01 \\
96X &       0.74 &     0.04&         -19.53 &      0.05 &      0.10 \\
89B &       0.83 &     0.06 &        -19.40 &      0.10 &     -0.02 \\
92A &       0.89 &     0.07 &        -19.31 &      0.11 &     -0.01 \\
86G &        1.06 &     0.07 &       -19.04 &      0.11 &     -0.12 \\
91bg &       0.94 &      0.21 &      -19.23 &      0.32  &     2.61 \\
\enddata
\tablenotetext{a}{$\delta M_{B}$ is the difference between the true
  value and that inferred using our correlation with the spectral
  index \RSiSu}
\end{deluxetable}



\subsubsection{\RCa}
\label{sec:rca}

Similar to \RSiSuS, \RCaS\  has a smaller dispersion due to 
noise, but a larger calibration error because of the coarse 
\snap\ binning. We thus only display \RCa\ results in this
section. \RCa\ calibration also suffers from the coarse
sampling of \snap\ spectrograph, as could be seen in the rightmost column
of Tables~\ref{tab:RCa2}--\ref{tab:RCa4}. Since we found no trend with
epoch, we have used the results for the slope without correction to
$t=0$ (Table~\ref{tab:RCaRCaSLinReg2}).

\begin{deluxetable}{lcccccc}
\tablecolumns{6}
\tablewidth{0pc}
\tablecaption{\RCa\ Case A $z=1.5$\label{tab:RCa2}}
\tablehead{
 \colhead{SN name}  &   \colhead{mean}   & \colhead{
   $\sigma_{\textrm{\RCa}}$} & \colhead{Inferred $M_{B}$} &
 \colhead{$\sigma_{M_{B}}$} & \colhead{$\delta M_{B}$}}
\startdata
91T &       0.88 &     0.03 &    -19.81 &     0.02 &      -0.20 \\
81B &       1.20 &     0.08 &   -19.53 &      0.07 &    -0.01  \\
98bu &      1.20 &     0.06 &   -19.53 &      0.05 &      0.04 \\
94D &       1.28 &     0.08 &   -19.46 &      0.07 &     -0.01  \\
96X &       1.42 &     0.08 &   -19.35 &      0.07 &     -0.08 \\
89B &       1.06 &     0.08 &   -19.65 &      0.07 &      0.23 \\
92A &       1.28 &     0.10 &   -19.47 &      0.08 &      0.15 \\
91bg &       2.43&     1.71 &   -18.48 &      1.46 &      1.86 \\
\enddata
\tablenotetext{a}{$\delta M_{B}$ is the difference between the true
  value and that inferred using our correlation with the spectral
  index \RCa.}
\end{deluxetable}

\begin{deluxetable}{lcccccc}
\tablecolumns{7}
\tablewidth{0pc}
\tablecaption{\RCa\ Case B $z=1.5$\label{tab:RCa4}}
\tablehead{
 \colhead{SN name}  &   \colhead{mean}   & \colhead{
   $\sigma_{\textrm{\RCa}}$} & \colhead{Inferred $M_{B}$} &
 \colhead{$\sigma_{M_{B}}$} & \colhead{$\delta M_{B}$}}
\startdata
91T &       0.88 &     0.03 &  -19.65 &      0.01 &      -0.36 \\
81B &       1.20 &     0.08 &  -19.49 &      0.04 &     -0.04 \\
98bu &      1.20 &     0.06 &  -19.49 &      0.03 &      0.00 \\
94D &       1.28 &     0.08 &  -19.46 &      0.04 &     -0.01 \\
96X &       1.42 &     0.08 &  -19.39 &      0.04 &     -0.04 \\
89B &       1.06 &     0.08 &  -19.56 &      0.04 &      0.14 \\
92A &       1.28 &     0.10 &  -19.46 &      0.05 &      0.14 \\
91bg &       2.43&     1.71 &  -18.92 &      0.8 &        2.29\\
\enddata
\tablenotetext{a}{$\delta M_{B}$ is the difference between the true
  value and that inferred using our correlation with the spectral
  index \RCa.}
\end{deluxetable}

As can be seen in these tables that summarize the luminosity measure
precision estimated for \RCa, the systematic error dominates, the
error due to the Poisson noise being always lower than $0.19$ blue
magnitudes.

\subsection{Evolution with $z$}
\label{sec:evol}

\snap\ wavelength coverage does not
allow the use either \RSi, or \RSiSu\ for $z=1.7$. On the other hand \RCa\ will
be measurable. JEDI would allow the measurement of \RSi\ at $z=1.7$,
but \RCa\  falls outside the range of its proposed IR spectrograph. Since the
exposure time has been 
calculated so 
that the number of photons per bin will remain almost constant,
the signal to noise will not be degraded too
much. The dispersion due to noise 
(Tables~\ref{tab:RCa1}--\ref{tab:RCa3} and Table~\ref{tab:snapressum})
remains small compared to the 
intrinsic dispersion of the \supernovae.  

\begin{deluxetable}{lcccccc}
\tablecolumns{7}
\tablewidth{0pc}
\tablecaption{\RCa\ Case A $z=1.7$\label{tab:RCa1}}
\tablehead{
 \colhead{SN name}  &   \colhead{mean}   & \colhead{
   $\sigma_{\textrm{\RCa}}$} & \colhead{Inferred $M_{B}$} &
 \colhead{$\sigma_{M_{B}}$} & \colhead{$\delta M_{B}$}
}
\startdata
91T &       0.80 &      0.03 &  -19.87 &      0.02 &      -0.13 \\
81B &       1.32 &     0.07 &  -19.43 &      0.06 &     -0.11 \\
98bu &      1.26 &     0.14 &  -19.48 &      0.12 &     -0.00 \\
94D &       1.25 &     0.09 &  -19.49 &      0.07 &      0.02\\
96X &       1.3 &     0.05 &  -19.45 &      0.04 &      0.02\\
89B &       1.04 &     0.08 &  -19.67 &      0.07 &      0.25\\
92A &       1.32 &     0.12 &  -19.42 &      0.1 &      0.1\\
91bg &       2.39 &    1.41 &  -18.51 &      1.2 &      1.89 \\
\enddata
\tablenotetext{a}{$\delta M_{B}$ is the difference between the true
  value and that inferred using our correlation with the spectral
  index \RCa.}
\end{deluxetable}

\begin{deluxetable}{lcccccc}
\tablecolumns{7}
\tablewidth{0pc}
\tablecaption{\RCa\ Case B $z=1.7$\label{tab:RCa3}}
\tablehead{
 \colhead{SN name}  &   \colhead{mean}   & \colhead{
   $\sigma_{\textrm{\RCa}}$} & \colhead{Inferred $M_{B}$}&
 \colhead{$\sigma_{M_{B}}$} & \colhead{$\delta M_{B}$}}
\startdata
91T &     0.80 &      0.03 &   -19.68 & 0.01 &     -0.33 \\
81B &     1.32 &     0.07 &    -19.43 & 0.03 &     -0.10 \\
98bu &    1.26 &     0.14 &    -19.47 & 0.06 &     -0.02 \\
94D &     1.25 &     0.09 &    -19.47 & 0.04 &     -0.00 \\
96X &     1.3 &     0.05 &     -19.45 & 0.02 &      0.02 \\
89B &     1.04 &     0.08 &    -19.57 & 0.04 &      0.15 \\
92A &     1.32 &     0.12 &    -19.43 & 0.06 &      0.11 \\
91bg&      2.39 &    1.41 &    -18.93 & 0.66 &      2.31\\
\enddata
\tablenotetext{a}{$\delta M_{B}$ is the difference between the true
  value and that inferred using our correlation with the spectral
  index \RSiSu.}
\end{deluxetable}

The \RCa\ blue magnitude accuracy remains dominated by the intrinsic
dispersion of the \supernovae corresponding to $\sigma_{M_{B}}\approx
  0.2$. This result can again 
be improved by using  the $80$ \supernovae  
expected. The blue magnitude precision would then be $\Delta M_{B}
\approx 0.02$, which is within the \snap\ cosmological requirements.

\begin{deluxetable}{llll}
\tablecolumns{7}
\tablewidth{0pc}
\tablecaption{Summary of Spectral Indicator Precision for
  \snap\label{tab:snapressum}} 
\tablehead{
 \colhead{Index}  &   \colhead{$\sigma_{M_{B}}$}   &
 \colhead{$\sigma_{M_{B}}$} &  \colhead{$\sigma_{M_{B}}$} \\
&   \colhead{(intrinsic)}   & \colhead{(noise)} & \colhead{ (total)} \\
}
\startdata
\RCa\ ($z=1.5$, no 91T, Case A)   &  0.16 &  0.14 &  0.21 \\
\RCa\ ($z=1.5$, with 91T, Case A) &  0.16 &  0.16 &  0.23 \\
\RCa\ ($z=1.5$, no 91T, Case B)   &  0.10 &  0.10 &  0.14 \\
\RCa\ ($z=1.5$, with 91T, Case B) &  0.10 &  0.19 &  0.21 \\
\RCa\ ($z=1.7$, no 91T, Case A)   &  0.16 &  0.16 &  0.23 \\
\RCa\ ($z=1.7$, with 91T, Case A) &  0.16 &  0.15 &  0.22 \\
\RCa\ ($z=1.7$, no 91T, Case B)   &  0.10 &  0.11 &  0.15 \\
\RCa\ ($z=1.7$, with 91T, Case B) &  0.10 &  0.18 &  0.20 \\
\RSi\ (no 91T, Case A)            &  0.12 &  0.24 &  0.27 \\
\RSi\ (with 91T, Case A)          &  0.12 &  1.31 &  1.32 \\
\RSi\ (no 91T, Case B)            &  0.10 &  0.14 &  0.17 \\
\RSi\ (with 91T, Case B)          &  0.10 &  0.65 &  0.66 \\
\RSiSu\ (no 91T, Case A)          &  0.05 &  0.11 &  0.12 \\
\RSiSu\ (with 91T, Case A)        &  0.05 &  0.16 &  0.17 \\
\RSiSu\ (no 91T, Case B)          &  0.04 &  0.09 &  0.10 \\
\RSiSu\ (with 91T, Case B)        &  0.04 &  0.16 &  0.16 \\
\enddata
\tablenotetext{a}{The methods used to estimate the error is described
  in the text. The method 
  is very conservative and double counts the systematic error.
Clearly \RSiSu\ without 91T-likes is the preferred method.}
\end{deluxetable}

\section{Conclusions}

We have presented a way to automatically measure the value of the
spectral indices \RCa\ and \RSi, which should be robust. We
have defined three new spectral indices \RCaS, \RSiSu, and \RSiSuS,
which are less sensitive to noise in the spectrum. We have shown that
for the limited set of Branch Normal supernovae where public spectra
are available near maximum light that there is a good correlation
between the spectral indices and the luminosity at peak. Within the
caveat of small size of our sample we found that there is not a
general trend in the spectral indices with epoch as long as the epoch
is close to maximum light. We have used synthetic spectral models to
show qualitatively that there does appear to be a deviation from
linearity in the spectral indices, particularly among the dim,
fast-declining \supernovae, such as SN~1991bg. This result is in
agreement with that of \citet{philetal99} who also found that there is a
deviation from linearity in peak magnitude with $\delta m_{15}$ as
$\delta m_{15}$ increases (see their Figure 8). It is also in
agreement with some theoretical hydrodynamical results which suggest
it may be difficult to produce a SN~Ia with such low nickel mass via a
deflagration (W.~Hillebrandt, private communication), although
\citet{HGFS99by02} are able to vary the nickel mass by varying the
density of the deflagration to detonation transition. 
It has been suggested that there is both a variation in
\SNeIa luminosity with galaxy type \citep{blue96,hametal01} and that
there are two populations of \SNeIa progenitors \citep{SB05,MDVP05}
which could be related to the deviation in the linearity of the
spectral indices.

Restricting ourselves to the sample of Branch Normal \supernovae we
find that the proposed JDEM mission SNAP could in fact use \RSiSu\ as a
complementary luminosity indicator. While the results may appear
somewhat circular in that we have restricted our calibrators to have a
small dispersion and then shown that the correlation based on that
result also recovers the small dispersion, the ability to use a
\emph{separate} observable from light-curve shape will provide for a
complementary measurement of the luminosity distance as a function of
redshift in the proposed JDEM wide-field high-redshift spaced based
surveys such as \snap\ and JEDI. The lightcurve based method of
template fitting will have a certain set of systematics that depends
on exactly how ``like versus like'' is chosen, whereas the spectral
index method will have a different set of systematics. In addition,
the spectral index methods are very weakly sensitive to both the
effects of dust and of the flux calibration of the spectrum, since
they are defined in a very narrow wavelength range. 

Our simulations show that the single spectrum obtained by
\snap\ for type identification will have sufficient signal to noise to
measure the spectral indices to the required precision. JEDI, which
will obtain multiple spectra for each supernova, would be even better
suited to using \RSiSu\ since they would have a larger sample of
spectra very close to maximum light.

As more spectra from nearby \supernovae with known peak luminosities
become available we should be able to refine our calibration, and thus
improve its accuracy for JDEM and other \supernovae
surveys. We note that our actual numerical results rely on both a very
small sample and also on the value of the absolute magnitudes assigned
by \citet{reindl05}, although since we quote results with both Case A
and B the sensitivity to the assigned absolute magnitudes can be
gauged somewhat. All available distance estimates (and hence absolute
magnitudes) to our sample will contain errors, and we have chosen to
use the results of \citet{reindl05} because they are derived in a
specific somewhat homogeneous manner. Thus, we regard this work as a
``proof of concept'' and the specific numerical values as our best
current estimate. Clearly when 300 Hubble-flow supernovae are
available from current ground-based surveys (i.e. in the next 3-5
years), this analysis should be recomputed. Once accurate values are
in hand, comparing absolute magnitude (determined from
light-curve shape) and \RSiSu\ determined at high redshift may provide a
useful measure of the evolutionary effects in the high-redshift
sample. We discuss this further in a forthcoming publication where we
study the formation of the \SiIIblue\ line.

\begin{acknowledgments}
  We thank Anne Ealet for information about the characteristics of SNAP
  and Peter Nugent for helpful discussions. This work was supported in
  part by NASA grants NAG5-3505 and NAG5-12127, NSF grants
  AST-0204771 and AST-0307323, and the Region Rhone-Alpes.  PHH was
  supported in part by the P\^ole 
  Scientifique de Mod\'elisation Num\'erique at ENS-Lyon.  This
  research used resources of:  the National Energy Research
  Scientific Computing Center (NERSC), which is supported by the
  Office of Science of the U.S.  Department of Energy under Contract
  No. DE-AC03-76SF00098; and the H\"ochstleistungs Rechenzentrum Nord
  (HLRN).  We thank both these institutions for a generous allocation
  of computer time.
\end{acknowledgments}

\clearpage

 

\end{document}